\documentclass[11pt]{article}

\usepackage[T1]{fontenc}
\usepackage[utf8]{inputenc}
\usepackage[a4paper,margin=2.45cm]{geometry}
\usepackage{amsmath,amssymb,bm,mathtools}
\usepackage{graphicx}
\usepackage{xcolor}
\usepackage{booktabs}
\usepackage{microtype}
\usepackage[numbers,square,sort&compress]{natbib}
\usepackage{caption}
\usepackage[colorlinks=true,linkcolor=blue,citecolor=blue,urlcolor=blue]{hyperref}
\usepackage{url}
\allowdisplaybreaks
\hypersetup{
  pdftitle={Block-Wise Variational Quantum Algorithms for Partial Differential Equations with Flux-Controlled Interface Penalties},
  pdfauthor={Jie Hangran}
}

\newcommand{\ee}{\mathrm{e}}
\newcommand{\ii}{\mathrm{i}}
\newcommand{\dd}{\mathrm{d}}
\newcommand{\ket}[1]{\left|#1\right\rangle}
\newcommand{\bra}[1]{\left\langle #1\right|}
\newcommand{\braket}[2]{\left\langle #1\middle|#2\right\rangle}

\newcommand{\jump}[1]{\left[\!\left[#1\right]\!\right]}

\newcommand{\Nh}{\widehat{\mathcal{N}}}
\newcommand{\Ah}{\widehat{A}}

\newcommand{\ResultPanelHeight}{0.235\textheight}
\providecommand{\doi}[1]{\href{https://doi.org/#1}{https://doi.org/#1}}

\begin{document}

\title{Block-Wise Variational Quantum Algorithms for PDEs with Interface Penalty Constraints}
\author{
Hangran Jie\textsuperscript{1}\thanks{\texttt{jiangran@hrbeu.edu.cn}} \qquad
Yuntao Cui\textsuperscript{1}\thanks{\texttt{YunTao\textunderscore Cui@outlook.com}} \qquad
Sunho Kim\textsuperscript{1}\thanks{Corresponding author: \texttt{kimsunho81@hrbeu.edu.cn}}\\[1em]
\small \textsuperscript{1}College of Mathematical Sciences, Harbin Engineering University, Harbin 150001, China
}

\maketitle

\begin{abstract}
Global variational quantum algorithm (VQA) frameworks for solving partial differential equations (PDEs) often rely on a single expressive ansatz over a uniform grid, which becomes structurally inefficient when solutions exhibit spatially heterogeneous complexity such as localized singularities or thin boundary layers. A localized nonsmooth feature can degrade the convergence of the entire global quantum representation, imposing unnecessary circuit depth and amplifying the risk of vanishing gradients in barren plateaus.

To overcome this limitation, we propose a block-wise VQA framework for PDEs characterized by spatially heterogeneous solution complexity. The computational domain is decomposed into locally represented quantum subproblems, where grid budgets and ansatz families are dynamically assigned based on a computable difficulty indicator. Artificial block interfaces are coupled through unified jump penalties for state values, derivatives, and physical fluxes, while adaptive reblocking tracks moving rough regions to maintain local accuracy without excessive global qubit overhead. The error analysis rigorously separates spatial discretization, ansatz expressivity, optimization convergence, finite-shot sampling noise, transfer errors from reblocking, and interface-coupling contributions. Reproducible residual-based simulations on representative elliptic, advection--diffusion, and Burgers problems demonstrate lower approximation errors and reduced peak local circuit width compared to global VQA approaches. These results substantiate block-wise quantum resource localization, confirming that high-fidelity PDE solutions can be achieved on near-term quantum devices without requiring a single globally expressive ansatz.
\end{abstract}


\section{Introduction}
\label{sec:introduction}

Variational quantum algorithms (VQAs) combine parametrized unitary circuits with classical optimizers and have emerged as the leading framework for leveraging noisy intermediate-scale quantum (NISQ) devices prior to the advent of fault-tolerant quantum computers~\cite{peruzzo2014variational,farhi2014qaoa,preskill2018nisq,cerezo2021variational,bharti2022noisy,tilly2022variational}.
However, their practical performance is constrained by multidimensional factors rather than a single metric.
Specifically, it depends simultaneously on the expressive power of the parametrized circuit and the trainability of the measurable cost function, which is intimately linked to gradient structures and the risk of barren plateaus~\cite{mcclean2018barren,cerezo2021cost,arrasmith2021effect}; the accumulation of shot noise from repeated circuit executions; state-preparation overhead; and, crucially, the extent to which the ansatz reflects the underlying symmetries or topological structure of the target physical observable~\cite{grant2019initialization,schuld2019evaluating,mitarai2018quantum,huang2020predicting}.

The numerical solution of partial differential equations (PDEs) represents a particularly important application domain for VQAs, as it necessitates the efficient manipulation of exponentially large Hilbert spaces arising from spatial discretization.
Existing quantum approaches include HHL-based quantum linear system algorithms, Hamiltonian-simulation methods, and their hybrid variants such as variational quantum linear solvers (VQLS) and variational PDE solvers (VQA-PDE)~\cite{harrow2009quantum,childs2017quantum,berry2014high,clader2013preconditioned,bravo2023variational,huang2019near,lubasch2020variational,sarma2024quantum}.
In standard VQA-PDE formulations, a discretized numerical solution is amplitude-encoded into a quantum state, a time-discrete PDE residual is mapped onto a measurable expectation-value loss function over a hypothesis space, and a classical optimizer updates the circuit parameters~\cite{lubasch2020variational,sarma2024quantum}.
Recent work on variational Poisson solvers has further emphasized that ansatz structural properties and double-factorization techniques for Hamiltonian or differential operators can critically impact the applicability of vanilla gradient descent and circuit-depth resources~\cite{ayoub2025poisson}.

Nevertheless, most existing formulations presuppose a single global grid and a single global ansatz.
This approach is computationally sound when the solution regularity is nearly uniform across the domain.
However, in many physical models, localized regions exhibit derivative jumps, thin boundary layers, moving fronts, or concentrated high-frequency components.
In such regimes, the convergence of global Fourier-based spectral methods is fundamentally compromised.
Fourier approximation is inherently governed by the regularity of the target function: solutions residing in finite Sobolev spaces yield only algebraic convergence, whereas analytic functions achieve geometric convergence~\cite{trefethen2000spectral,boyd2001chebyshev,canuto2006spectral,shen2011spectral}.
Consequently, a single localized singularity can dictate the convergence rate of the entire global Fourier series, forcing the global ansatz to suffer from unnecessary representational overhead and an elevated risk of vanishing gradients, even if the remaining domain is perfectly smooth.

To address this structural inefficiency, we develop a block-wise VQA-PDE framework.
Our methodology decomposes the computational domain into local blocks, assigning an independent spatial-discretization budget (grid density) and a problem-tailored ansatz to each block.
The quantum circuits of individual blocks are weakly coupled through physical boundary conditions and interface residual terms.
Specifically, smooth blocks employ shallow-depth Fourier or Quantum Fourier Transform (QFT)-based representations to ensure global convergence, while blocks containing persistent high-frequency noise or nonsmooth components utilize layered ansatzes or deeper circuits to absorb local complexity.
We clarify that the selection of a layered ansatz for rough blocks constitutes an empirical design decision based on algorithmic efficiency and observational evidence, rather than a universal theorem guaranteeing that every rough function admits a prescribed uniform logical axiom (ULA) convergence rate.

The principal technical challenge in this process lies in handling discontinuities at artificial interfaces.
Independently evolved block states need not naturally satisfy continuity in either function values or PDE fluxes across block boundaries.
To mathematically rectify this, we borrow external principles established in Discontinuous Galerkin (DG), interior-penalty, and domain-decomposition methods.
These methodologies inherently couple discontinuous local trial functions (broken Sobolev spaces) through jump penalties, numerical fluxes, and stabilization terms at interfaces~\cite{reed1973triangular,bassi1997high,cockburn1998local,cockburn2001runge,arnold2002unified,riviere2008discontinuous,hesthaven2008nodal,dipietro2012mathematical,nitsche1971,dolean2015domain}.
While DG methods are classical numerical tools, their core principle---coupling discontinuous local representations through interface penalties---maps naturally onto the problem of coupling independent quantum circuit blocks with overlapping or adjacent support.
We employ this principle solely to motivate and analyze the interface residual within the quantum-circuit context.
We explicitly note that merely transplanting this mathematical principle to a quantum representation does not by itself prove that a block-wise approach always yields superior accuracy or lower resource costs compared to a global VQA.

Furthermore, the present formulation introduces five essential distinctions that are indispensable for a defensible and rigorous error analysis:
\begin{enumerate}
    \item Separation of boundary conditions: Boundary conditions are not internalized within the local PDE loss function but are explicitly represented as an independent residual term.
    \item Unified interface orientation: Jump terms at interfaces are defined with respect to a common orientation, preemptively eliminating the sign ambiguity that arises from mixing outward normals of adjacent blocks.
    \item Explicit residual-stability hypothesis: Connecting interface penalty terms to the actual solution error of the quantum state requires an explicit residual-stability assumption.
    \item Decoupling of error sources: The representational limits of the ansatz, the convergence of classical optimization, and the statistical noise contribution from finite-shot measurements are rigorously defined as separate terms; the variance of the loss estimator must not be directly equated with the physical solution error.
    \item Adaptive reblocking: Transfer errors introduced during grid adaptation are triggered from the reconstructed approximation of the current circuit rather than from the inaccessible true solution.
\end{enumerate}

Built upon this rigorous foundation, the specific contributions of this work are as follows:
(i) We formulate local amplitude-encoded states incorporating physical boundary residuals and PDE-dependent interface losses, and define an executable procedure for monitoring, domain partitioning, ansatz selection, and power-of-two quantum register resource allocation.
(ii) We derive error bounds for blockwise spatial approximation and Fourier spectral estimates, ensuring that the correct local-length scaling of each block is properly reflected.
(iii) Under a stated stability hypothesis, we mathematically present conditional bounds for interface coupling, optimization convergence, and finite-shot measurement contributions to the total error.
(iv) We describe how low-frequency coefficient registers, circuit modules implementing local differential operators, and phase-sensitive endpoint measurements map onto an actual quantum-circuit-level implementation.
Our framework demonstrates that spatially adaptive quantum representations can achieve high-fidelity solutions while mitigating the exponential resource scaling that typically hinders global VQA approaches on near-term devices.

To maintain theoretical rigor, the empirical numerical evidence presented in this paper is kept strictly separate from these theoretical error propositions.
Specifically, classical reduced-basis fitting, exact-expectation loss minimization, finite-shot circuit simulation, and actual hardware execution constitute distinct validation levels and must never be conflated in academic interpretation.

The remainder of this paper is organized as follows.
Section~\ref{sec:preliminaries} concisely reviews the established external results required for the analysis and implementation of this work.
Section~\ref{sec:block_method_error} mathematically introduces the proposed block-wise methodology and derives its conditional error controls.
Section~\ref{sec:numerics} reports numerical test results, and finally, Sec.~\ref{sec:conclusion} concludes the paper.

\section{Preliminaries}
\label{sec:preliminaries}

This section contains background results from the literature. The proposed monitor, block construction, resource rule, boundary loss, interface penalty, and error decomposition are deferred to Sec.~\ref{sec:block_method_error}.

\subsection{Baseline VQA-PDE formulation}
\label{subsec:vqa_pde_baseline}

Consider a first-order-in-time PDE with a boundary operator $\mathcal B$,
\begin{equation}
  \partial_tu(x,t)=\mathcal N[u](x,t),
  \qquad
  \mathcal B u=g
  \quad\text{on }\partial\Omega.
  \label{eq:pde_general}
\end{equation}
The spatially discretized solution on $N=2^n$ grid points can be encoded as an unnormalized state~\cite{lubasch2020variational,sarma2024quantum}
\begin{equation}
  \ket{u(t)}=\alpha(t)\ket{\psi(\bm\theta(t))},
  \qquad
  \ket{\psi(\bm\theta)}=U(\bm\theta)\ket{0}^{\otimes n}.
  \label{eq:vqa_state}
\end{equation}

For a possibly nonlinear operator $\mathcal N$, a backward-Euler step is represented by the residual
\begin{equation}
  \ket{r^{m+1}}
  =\ket{u^{m+1}}-\ket{u^m}
  -\tau\ket{\mathcal N[u^{m+1}]},
  \label{eq:nonlinear_backward_residual}
\end{equation}
where nonlinear terms may require diagonal multiplication operators and intermediate variational states~\cite{lubasch2020variational,sarma2024quantum}. Only when $\mathcal N$ is represented by a linear operator $\Nh$ does Eq.~\eqref{eq:nonlinear_backward_residual} reduce to
\begin{equation}
  (I-\tau\Nh)\ket{u^{m+1}}-\ket{u^m}.
  \label{eq:linear_backward_residual}
\end{equation}
The corresponding squared residual cost is
\begin{equation}
\begin{aligned}
  C_u(\alpha,\bm\theta)
  =\Big\|
  &(I-\tau\Nh)\alpha\ket{\psi(\bm\theta)}\\
  &-\widetilde\alpha\ket{\widetilde\psi}
  \Big\|^2,
\end{aligned}
  \label{eq:baseline_cost}
\end{equation}
where $\widetilde\alpha\ket{\widetilde\psi}=\ket{u^m}$ denotes the known state from the preceding time level. Thus the tilde marks known previous-step data, not an additional ansatz.

On a periodic grid of length $L$, centered first- and second-difference operators can be expressed by an adder operator $\Ah$~\cite{lubasch2020variational,sarma2024quantum,leveque2007finite,strikwerda2004finite}:
\begin{equation}
\begin{aligned}
  \widehat\partial_x
  &=\frac{2^{n-1}}{L}(\Ah-\Ah^\dagger),\\
  \widehat\partial_{xx}
  &=\frac{4^n}{L^2}(\Ah+\Ah^\dagger-2I).
\end{aligned}
  \label{eq:adder_derivatives}
\end{equation}
These expressions contain periodic wrap-around. They cannot be applied unchanged to a nonperiodic local block; boundary rows must instead be replaced by one-sided stencils, ghost-value closures, or an equivalent low-rank boundary correction~\cite{leveque2007finite,strikwerda2004finite}.

\subsection{Ansatz families and Fourier approximation}
\label{subsec:ansatz_fourier_prelim}

Fourier/ZGR-QFT constructions load a compact set of Fourier coefficients and use a QFT-type transformation to obtain physical-grid amplitudes~\cite{grover2002creating,mottonen2005transformation,shende2006synthesis,plesch2011quantum,moosa2023linear,sarma2024quantum}. Layered circuits such as the ULA provide a more flexible variational family and have been used for jagged or nonsmooth states in VQA-PDE studies~\cite{sarma2024quantum}. The latter observation motivates an architecture choice, but it does not supply an architecture-independent ULA error rate.

For a periodic function on the reference interval,
\begin{equation}
  v(\xi)=\sum_{k\in\mathbb Z}\widehat v_k\ee^{2\pi\ii k\xi},
  \qquad
  P_Mv=\sum_{|k|\le M}\widehat v_k\ee^{2\pi\ii k\xi},
  \label{eq:fourier_projection}
\end{equation}
Parseval's identity gives
\begin{equation}
  \|v-P_Mv\|_{L^2(0,1)}^2
  =\sum_{|k|>M}|\widehat v_k|^2.
  \label{eq:parseval_tail}
\end{equation}
Standard spectral approximation estimates imply~\cite{trefethen2000spectral,boyd2001chebyshev,canuto2006spectral,shen2011spectral}
\begin{equation}
\begin{aligned}
  \|v-P_Mv\|_{L^2}
  &\le C_sM^{-s}\|v\|_{H^s},\\
  \|v-P_Mv\|_{L^2}
  &\le A\ee^{-\sigma M}
  \quad\text{for analytic }v.
\end{aligned}
  \label{eq:fourier_tail_estimates}
\end{equation}
For analytic $v$, derivative-sensitive estimates have the form
\begin{equation}
  \|\partial_\xi^r(v-P_Mv)\|_{L^2}
  \le A_rM^r\ee^{-\sigma M},
  \qquad r=0,1,2,
  \label{eq:derivative_fourier_tail}
\end{equation}
with constants depending on the analyticity region~\cite{trefethen2000spectral,shen2011spectral}.

\subsection{DG notation and PDE-dependent interface fluxes}
\label{subsec:dg_flux_prelim}

Let $\Gamma_i=\{x_i\}$ separate $I_i$ and $I_{i+1}$, and choose one fixed interface orientation $n_i$ pointing from $I_i$ to $I_{i+1}$. The same direction is used on both traces. We define
\begin{equation}
\begin{aligned}
  \jump{v}_{\Gamma_i}
  &=v_i(x_i^-)-v_{i+1}(x_i^+),\\
  \jump{\partial_{n_i}v}_{\Gamma_i}
  &=\nabla v_i(x_i^-)\!\cdot n_i
    -\nabla v_{i+1}(x_i^+)\!\cdot n_i.
\end{aligned}
  \label{eq:jump_notation}
\end{equation}
If separate outward normals $n_i^-=-n_i^+$ are used instead, conservation is expressed by the sum of the two outward normal fluxes, not their difference. This orientation convention is standard in DG and interface formulations~\cite{arnold2002unified,riviere2008discontinuous,hesthaven2008nodal,dipietro2012mathematical}.

For a convection--diffusion equation~\cite{leveque2007finite,strikwerda2004finite,quarteroni2007numerical}
\begin{equation}
  \partial_tu+\partial_xf(u)=\nu\partial_{xx}u,
  \label{eq:convection_diffusion}
\end{equation}
the physical flux in the positive $x$ direction is
\begin{equation}
  F(u,u_x)=f(u)-\nu u_x.
  \label{eq:physical_flux_general}
\end{equation}
For viscous Burgers, $f(u)=u^2/2$. A general transmission problem prescribes target jumps
\begin{equation}
  \jump{u}_{\Gamma_i}=g_{u,i},
  \qquad
  \jump{F(u,u_x)}_{\Gamma_i}=g_{F,i}.
  \label{eq:exact_interface_conditions}
\end{equation}
At an artificial interface of a smooth problem without an interface source, $g_{u,i}=g_{F,i}=0$. A manufactured derivative jump or a material interface can instead require $g_{F,i}\ne0$. Periodic problems also contain a closing interface connecting the right trace at $b$ to the left trace at $a$.

\subsection{Trace and residual-stability estimates}
\label{subsec:trace_error_prelim}

For $e_i\in H^1(I_i)$, the one-dimensional trace inequality gives~\cite{arnold2002unified,riviere2008discontinuous,ern2004theory,quarteroni2007numerical}
\begin{equation}
\begin{aligned}
  |e_i(x_i^-)|^2
  \le C\bigl(&\ell_i^{-1}\|e_i\|_{L^2(I_i)}^2\\
  &+\ell_i\|\partial_xe_i\|_{L^2(I_i)}^2\bigr).
\end{aligned}
  \label{eq:trace_value}
\end{equation}
For $e_i\in H^2(I_i)$,
\begin{equation}
\begin{aligned}
  |\partial_xe_i(x_i^-)|^2
  \le C\bigl(&\ell_i^{-1}\|\partial_xe_i\|_{L^2(I_i)}^2\\
  &+\ell_i\|\partial_{xx}e_i\|_{L^2(I_i)}^2\bigr).
\end{aligned}
  \label{eq:trace_derivative}
\end{equation}

DG and residual analyses for well-posed coercive problems control a broken-space error by interior, physical-boundary, and interface residuals~\cite{arnold2002unified,riviere2008discontinuous,dipietro2012mathematical,ern2004theory}. In schematic notation, the external stability principle used later is
\begin{equation}
  \|e\|_X
  \le C_{\mathrm{stab}}
  \left(
  \|\mathcal R_\Omega\|_{Y'}
  +\|\mathcal R_{\partial\Omega}\|_{B}
  +\|\mathcal R_\Gamma\|_{G}
  \right).
  \label{eq:background_residual_stability}
\end{equation}
The norm $X$, the residual norms, and $C_{\mathrm{stab}}$ depend on the PDE and discretization. Equation~\eqref{eq:background_residual_stability} is not automatic for every nonlinear PDE; the block-wise result in Sec.~\ref{subsec:error_factors} is explicitly conditional on an estimate of this form.

\subsection{Adaptive discretization, truncation, and measurement estimates}
\label{subsec:adaptive_measurement_prelim}

Adaptive mesh methods often use a positive monitor $m(x)$ and an equidistribution principle~\cite{tang2005moving,budd2009adaptivity,quarteroni2007numerical},
\begin{equation}
  \int_{x_{i-1}}^{x_i}m(x)\,\dd x
  \approx\frac1K\int_a^b m(x)\,\dd x.
  \label{eq:prelim_equidistribution}
\end{equation}
For a $q$th-order spatial approximation on $I_i$ with $N_i$ points, a standard local model is~\cite{leveque2007finite,strikwerda2004finite,quarteroni2007numerical}
\begin{equation}
  \epsilon_{x,i}
  \lesssim C_i^{(q)}\ell_i^{q+1/2}N_i^{-q},
  \label{eq:prelim_local_truncation}
\end{equation}
where $C_i^{(q)}$ contains the appropriate local higher-derivative norm and scheme constant. Combining block contributions in an $L^2$ sense yields
\begin{equation}
  E_x^2
  \lesssim\sum_i(C_i^{(q)})^2\ell_i^{2q+1}N_i^{-2q}.
  \label{eq:prelim_spatial_sum}
\end{equation}

Finally, if a measured quantity is a weighted sum $\widehat O=\sum_r c_r\widehat O_r$ and each term is estimated independently with $M_r$ shots, then~\cite{zhu2024optimizing,nakaji2023measurement}
\begin{equation}
  \operatorname{Var}(\widehat O)
  =\sum_r\frac{c_r^2\sigma_r^2}{M_r}.
  \label{eq:prelim_shot_variance}
\end{equation}
This is an estimator-variance formula. Converting it into a solution error requires an additional stability relation, which is supplied conditionally in Sec.~\ref{subsec:error_factors}.

\section{Proposed Block-Wise VQA-PDE Method}
\label{sec:block_method_error}

All formulas in this section are either definitions of the proposed method, short derivations from Sec.~\ref{sec:preliminaries}, or conditional consequences of the residual-stability estimate in Eq.~\eqref{eq:background_residual_stability}.

To separate the error sources, introduce a sequence of comparison states. Let $u^{(0)}=u$ be the exact PDE solution, $u^{(1)}$ the exact time-discrete solution, $u^{(2)}$ its spatially discrete projection, $u^{(3)}$ the best block-ansatz approximation satisfying the exact boundary and transmission conditions, $u^{(4)}$ the exact minimizer of the penalized block loss, $u^{(5)}$ the output of a finite-accuracy classical optimizer using exact expectations, and $u^{(6)}$ the output obtained with finite-shot estimates. Define
\begin{equation}
\begin{aligned}
  E_t&=\|u^{(0)}-u^{(1)}\|_X,
  &E_x&=\|u^{(1)}-u^{(2)}\|_X,\\
  E_{\mathrm{ans}}&=\|u^{(2)}-u^{(3)}\|_X,
  &\boldsymbol E_{\mathrm{int}}&=\|u^{(3)}-u^{(4)}\|_X,\\
  E_{\mathrm{opt}}&=\|u^{(4)}-u^{(5)}\|_X,
  &E_{\mathrm{shot}}&=\|u^{(5)}-u^{(6)}\|_X.
\end{aligned}
  \label{eq:error_stage_definitions}
\end{equation}
The triangle inequality therefore gives the proposed accounting bound
\begin{equation}
  E_{\mathrm{tot}}^{\mathrm{blk}}
  \le E_t+E_x+E_{\mathrm{ans}}+E_{\mathrm{opt}}+E_{\mathrm{shot}}
  +\boldsymbol E_{\mathrm{int}}.
  \label{eq:block_error_decomp_prelim}
\end{equation}
For a monolithic global method the interface stage is absent, so $u^{(3)}=u^{(4)}$ and $\boldsymbol E_{\mathrm{int}}=0$. Equations~\eqref{eq:error_stage_definitions} and~\eqref{eq:block_error_decomp_prelim} are definitions and a direct triangle-inequality derivation; they are not claims that the six contributions are statistically independent.

\subsection{Domain decomposition and local representation}
\label{subsec:domain_local_representation}

Let $\Omega=(a,b)$ and
\begin{equation}
\begin{aligned}
  \overline\Omega&=\bigcup_{i=1}^K\overline I_i,
  &I_i&=(x_{i-1},x_i),\\
  a&=x_0<x_1<\cdots<x_K=b.
\end{aligned}
  \label{eq:block_partition_new}
\end{equation}
Define $\ell_i=x_i-x_{i-1}$ and $\Gamma_i=\{x_i\}$ for $i=1,\ldots,K-1$. For periodic boundary conditions, the interface set is augmented by $\Gamma_{\mathrm{per}}$, which couples the trace at $b^-$ to the trace at $a^+$.

Block $i$ uses $n_i$ qubits and $N_i=2^{n_i}$ amplitudes. If $\mathbf u_i=(u_{i,0},\ldots,u_{i,N_i-1})^T$ is real-valued sampled data, define
\begin{equation}
\begin{aligned}
  \alpha_i&=\|\mathbf u_i\|_2,
  &\ket{\psi_i}
  &=\frac1{\alpha_i}\sum_{j=0}^{N_i-1}u_{i,j}\ket j,\\
  \ket{u_i}
  &=\alpha_iU_i(\bm\theta_i)\ket0^{\otimes n_i}.
\end{aligned}
  \label{eq:block_state_new}
\end{equation}
Thus $\alpha_i$ restores the physical amplitude scale. Because the blocks are executed separately,
\begin{equation}
  n_{\mathrm{peak}}=\max_i n_i,
  \qquad
  N_{\mathrm{used}}=\sum_i2^{n_i}.
  \label{eq:Ntot_block_new}
\end{equation}
The first quantity is peak circuit width; the second is a total grid budget. A reduction in $n_{\mathrm{peak}}$ does not by itself prove a reduction in total parameters, gates, circuit calls, or shots.

On block $I_i$, the physical grid spacing is $h_i=\ell_i/(N_i-1)$ when both endpoint traces are represented. Local derivative matrices use $h_i$ and replace periodic wrap-around rows by boundary/interface closures. The precise closure is part of the discretization and must be reported with the numerical experiment.

\subsection{Difficulty indicator, partition, and resource allocation}
\label{subsec:allocation}

Let $v$ be the available profile: $v=u_0$ for an initial fixed partition or $v=u_h(\cdot,t_m)$ for adaptive reblocking. On a window $W_j$ centered at grid point $x_j$, let $\widehat v_{j,k}$ denote the windowed discrete Fourier coefficients. We define the local high-frequency fraction by
\begin{equation}
  \tau_{\mathrm{loc}}(x_j)
  =\left(
  \frac{\sum_{k\in\mathcal K_{\mathrm{high}}}|\widehat v_{j,k}|^2}
       {\sum_k|\widehat v_{j,k}|^2+\varepsilon}
  \right)^{1/2}.
  \label{eq:local_tail_definition}
\end{equation}
The window size, taper, and high-frequency set $\mathcal K_{\mathrm{high}}$ are algorithmic parameters. Derivatives are computed from the same filtered snapshot so that noise in $v_h$ is not differentiated without regularization.

Choose $c_1,c_2,c_3\ge0$ with $c_1+c_2+c_3=1$. The proposed dimensionless monitor is
\begin{equation}
\begin{aligned}
  \eta(x)
  &=c_1\frac{|v_x(x)|}{\|v_x\|_{L^\infty}+\varepsilon}
  +c_2\frac{|v_{xx}(x)|}{\|v_{xx}\|_{L^\infty}+\varepsilon}\\
  &\quad+c_3\frac{\tau_{\mathrm{loc}}(x)}
  {\|\tau_{\mathrm{loc}}\|_{L^\infty}+\varepsilon}.
\end{aligned}
  \label{eq:difficulty_indicator_new}
\end{equation}
Hence $0\le\eta\lesssim1$. This is a proposed monitor, not a PDE identity and not the exact truncation constant in Eq.~\eqref{eq:prelim_local_truncation}. Other monitor choices are possible; the present combination is a practical realization chosen to detect precisely the gradient, curvature, and local high-frequency features that challenge the Fourier/ZGR-type representation used on smooth blocks.

The partition is constructed deterministically as follows.
\begin{enumerate}
  \item Mark $\mathcal R_{\mathrm{rough}}=\{x:\eta(x)>\theta_{\mathrm{rough}}\}$ and compute its connected components.
  \item Expand every marked component by a prescribed guard width $w_g$ so that a rough feature is not placed directly on an interface.
  \item Use the expanded component endpoints as candidate interfaces. Merge any interval shorter than $\ell_{\min}$ with the neighboring interval having the closest monitor average.
  \item If the number of blocks exceeds $K_{\max}$, merge the adjacent pair with the smallest combined monitor mass. If additional smooth blocks are desired, place them by the equidistribution rule in Eq.~\eqref{eq:prelim_equidistribution}.
\end{enumerate}
The parameters $\theta_{\mathrm{rough}},w_g,\ell_{\min},K_{\max}$ must be fixed before evaluating a reported test.

The theoretical constant $C_i^{(q)}$ in Eq.~\eqref{eq:prelim_local_truncation} depends on higher derivatives and the chosen spatial scheme. We do not identify it with $\eta$. Instead, define the empirical allocation weight
\begin{equation}
\begin{aligned}
  \widehat C_i
  &=C_{\min}\left(1+\gamma_C\overline\eta_i\right),\\
  \overline\eta_i
  &=\left(\frac1{\ell_i}\int_{I_i}\eta(x)^2\,\dd x\right)^{1/2},
\end{aligned}
  \label{eq:Ci_indicator_new}
\end{equation}
where $C_{\min}>0$ and $\gamma_C\ge0$ are calibration parameters. Equation~\eqref{eq:Ci_indicator_new} is a computable surrogate weight. The allocation below is optimal only for the resulting estimated model,
\begin{equation}
  \mathcal M_{\mathrm{est}}
  =\sum_{i=1}^K\widehat C_i^2\ell_i^{2q+1}N_i^{-2q}.
  \label{eq:allocation_objective}
\end{equation}

For a continuous budget $B_N$ with $\sum_iN_i=B_N$, a Lagrange-multiplier calculation gives
\begin{equation}
  -2q\widehat C_i^2\ell_i^{2q+1}N_i^{-2q-1}+\mu=0,
  \label{eq:allocation_stationary}
\end{equation}
and therefore
\begin{equation}
  N_i^{\mathrm{cont}}
  =B_N\frac{\ell_i\widehat C_i^{2/(2q+1)}}
  {\sum_j\ell_j\widehat C_j^{2/(2q+1)}}.
  \label{eq:continuous_allocation_new}
\end{equation}
This derives the powers of $\ell_i$ and $\widehat C_i$ rather than assuming them.

The quantum constraint is $\sum_i2^{n_i}\le B_N$, which may not admit equality. Initialize
\begin{equation}
  n_i=\max\left\{n_{\min},
  \left\lfloor\log_2N_i^{\mathrm{cont}}\right\rfloor\right\}.
  \label{eq:raw_qubits_new}
\end{equation}
If this initialization exceeds $B_N$, remove qubits from the block with the smallest estimated error increase per removed grid point, subject to $n_i\ge n_{\min}$. Then add one qubit at a time while the budget permits. For a candidate doubling $N_i\mapsto2N_i$, use
\begin{equation}
  \rho_i
  =\frac{\widehat C_i^2\ell_i^{2q+1}
  [N_i^{-2q}-(2N_i)^{-2q}]}{N_i}
  \label{eq:marginal_gain_new}
\end{equation}
as the estimated error reduction per additional grid point, and select the feasible block with largest $\rho_i$. The procedure stops when no doubling fits inside $B_N$.

For ansatz selection define
\begin{equation}
  \eta_i^{\max}=\max_{x\in I_i}\eta(x),
  \qquad
  \tau_i^{\mathrm{avg}}
  =\frac1{\ell_i}\int_{I_i}\tau_{\mathrm{loc}}(x)\,\dd x.
  \label{eq:block_indicators_new}
\end{equation}
The executable rule is
\begin{equation}
  \mathcal A_i=
  \begin{cases}
    \text{ZGR-QFT},&
    \eta_i^{\max}\le\theta_s
    \text{ and }\tau_i^{\mathrm{avg}}\le\theta_{\mathrm{tail}},\\
    \text{ULA},&\text{otherwise}.
  \end{cases}
  \label{eq:ansatz_selection_new}
\end{equation}
For a ZGR-QFT block,
\begin{equation}
  M_i=\min\left\{M_{\max},\left\lfloor\frac{N_i-1}{2}\right\rfloor\right\},
  \label{eq:Mi_rule_new}
\end{equation}
whereas a ULA block uses the empirical depth rule
\begin{equation}
  d_i=d_{\min}+\left\lceil\gamma_d\eta_i^{\max}\right\rceil.
  \label{eq:depth_rule_new}
\end{equation}
The ULA rule must be validated by depth and initialization sweeps; it is not inferred from the Fourier theorem.

\subsection{Boundary- and interface-penalized loss}
\label{subsec:interface_loss}

Let $\mathcal L_i$ be the mean squared local PDE residual on $I_i$. To approximate a domain-integrated loss without over-weighting small blocks, set
\begin{equation}
  \mathcal L_{\mathrm{PDE}}
  =\sum_{i=1}^K w_i\mathcal L_i,
  \qquad
  w_i=\frac{\ell_i}{|\Omega|}.
  \label{eq:LPDE_new}
\end{equation}
If $\mathcal L_i$ is already an unnormalized quadrature sum, the quadrature weights replace $w_i$; the convention must be stated consistently.

Let $U_*>0$, $D_*=U_*/L_*>0$, and $F_*>0$ be characteristic value, derivative, and flux scales. Define the normalized target residuals
\begin{equation}
\begin{aligned}
  \rho_{u,i}
  &=\frac{\jump{u_h}_{\Gamma_i}-g_{u,i}}{U_*},\\
  \rho_{n,i}
  &=\frac{\jump{\partial_{n_i}u_h}_{\Gamma_i}-g_{n,i}}{D_*},\\
  \rho_{F,i}
  &=\frac{\jump{F(u_h,\partial_xu_h)}_{\Gamma_i}-g_{F,i}}{F_*}.
\end{aligned}
  \label{eq:normalized_interface_residuals}
\end{equation}
The proposed interface loss is
\begin{equation}
  \mathcal J_\Gamma
  =\sum_{\Gamma_i\in\mathcal G_\Gamma}
  \left(
  \lambda_u|\rho_{u,i}|^2
  +\lambda_n|\rho_{n,i}|^2
  +\lambda_F|\rho_{F,i}|^2
  \right),
  \label{eq:JGamma_new}
\end{equation}
where $\mathcal G_\Gamma$ includes the periodic closing interface when applicable. The derivative penalty is optional. For constant-coefficient convection--diffusion, value and physical-flux control can make it redundant; the active terms must be specified for each PDE.

Physical boundary conditions are imposed separately. With boundary trace operators $B_D$, $B_N$, and $B_R$ for Dirichlet, normal-flux/Neumann, and Robin data, respectively, define
\begin{equation}
\begin{aligned}
  \mathcal J_{\partial\Omega}
  =&\ \lambda_D\|B_Du_h-g_D\|^2
  +\lambda_N\|B_NF_h-g_N\|^2\\
  &+\lambda_R\|B_Ru_h-g_R\|^2.
\end{aligned}
  \label{eq:Jboundary_new}
\end{equation}
Only the terms corresponding to the prescribed boundary type are retained. A strongly encoded boundary condition may be omitted from Eq.~\eqref{eq:Jboundary_new}, but that choice must be explicit. The total loss is
\begin{equation}
  \mathcal L_{\mathrm{tot}}
  =\mathcal L_{\mathrm{PDE}}
  +\mathcal J_\Gamma
  +\mathcal J_{\partial\Omega}.
  \label{eq:Ltotal_new}
\end{equation}
The interface penalty changes the variational objective and therefore can move the trained parameters away from the minimizer of the unpenalized local residual.  Consequently, increasing $\lambda_u$, $\lambda_n$, or $\lambda_F$ can reduce $\boldsymbol E_{\mathrm{int}}$ while slightly increasing the field error through an effective constrained-approximation or optimization bias.  This tradeoff does not alter the formal best-approximation definition of $E_{\mathrm{ans}}$ above, but in finite-depth, finite-budget computations it changes the realized ansatz parameters.  Experiment~2 therefore reports a penalty sweep rather than assuming that a larger penalty is always better.

\subsection{Error factors and conditional controls}
\label{subsec:error_factors}

Applying the standard estimate in Eq.~\eqref{eq:prelim_local_truncation} block by block gives
\begin{equation}
  (E_x^{\mathrm{blk}})^2
  \lesssim\sum_{i=1}^K(C_i^{(q)})^2
  \ell_i^{2q+1}N_i^{-2q}.
  \label{eq:spatial_factor_sum_new}
\end{equation}
The allocation rule minimizes the corresponding estimated model with $C_i^{(q)}$ replaced by $\widehat C_i$; it is optimal for that model, not an unconditional proof that $E_x^{\mathrm{blk}}<E_x^{\mathrm{uni}}$.

For a ZGR block, map $I_i$ to $[0,1]$ by $v_i(\xi)=u(x_{i-1}+\ell_i\xi)$. Since $\partial_x^r=\ell_i^{-r}\partial_\xi^r$ and $\dd x=\ell_i\dd\xi$,
\begin{equation}
  \|\partial_x^r(u-P_{M_i}u)\|_{L^2(I_i)}
  =\ell_i^{1/2-r}
  \|\partial_\xi^r(v_i-P_{M_i}v_i)\|_{L^2(0,1)}.
  \label{eq:local_scaling_identity}
\end{equation}
Consequently,
\begin{equation}
  E_{\mathrm{ans},i}^{\mathrm{ZGR}}
  \le \ell_i^{1/2}C_{s_i}M_i^{-s_i}
  \|v_i\|_{H^{s_i}(0,1)},
  \label{eq:ansatz_factor_algebraic_new}
\end{equation}
and for analytic $v_i$,
\begin{equation}
  \|\partial_x^r(u-P_{M_i}u)\|_{L^2(I_i)}
  \le\ell_i^{1/2-r}A_{i,r}M_i^r\ee^{-\sigma_iM_i}.
  \label{eq:ansatz_factor_geometric_new}
\end{equation}
The factors $\ell_i^{1/2-r}$ are essential when a flux contains physical derivatives.

For a ULA block, the honest architecture-dependent quantity is the best-approximation error
\begin{equation}
  E_{\mathrm{ans},i}^{\mathrm{ULA}}(d_i,n_i)
  =\inf_{\bm\theta_i,\alpha_i}
  \|u_i-\alpha_iU_i^{\mathrm{ULA}}(\bm\theta_i)\ket0^{\otimes n_i}\|_X.
  \label{eq:ula_best_approximation_new}
\end{equation}
No universal rate in $d_i$ is asserted. Its value must be estimated numerically for the stated circuit, initialization, and optimizer.

We next connect the computable interface loss to $\boldsymbol E_{\mathrm{int}}$. Assume that the PDE and the chosen time-discrete block formulation satisfy Eq.~\eqref{eq:background_residual_stability} in a neighborhood of the exact solution. For the comparison $u^{(3)}-u^{(4)}$, also assume that changes in the interior and physical-boundary residuals are either zero or assigned to the spatial, ansatz, and optimization stages, so that the restricted estimate $\|u^{(3)}-u^{(4)}\|_X\le C_\Gamma\|\mathcal R_\Gamma\|_G$ holds. This additional statement is the precise stability condition needed to isolate an interface contribution. Let
\begin{equation}
  R_\Gamma^2
  =\sum_{\Gamma_i\in\mathcal G_\Gamma}
  (|\rho_{u,i}|^2+|\rho_{n,i}|^2+|\rho_{F,i}|^2),
  \label{eq:interface_residual_norm_new}
\end{equation}
with inactive components omitted, and let $\lambda_{\min}$ be the smallest active penalty weight. Then Eq.~\eqref{eq:JGamma_new} gives $R_\Gamma\le\lambda_{\min}^{-1/2}\mathcal J_\Gamma^{1/2}$. The stability estimate therefore yields the conditional interface control
\begin{equation}
  \boldsymbol E_{\mathrm{int}}
  \le C_\Gamma R_\Gamma
  \le\frac{C_\Gamma}{\sqrt{\lambda_{\min}}}
  \mathcal J_\Gamma^{1/2}.
  \label{eq:Eint_penalty_control_new}
\end{equation}
Equation~\eqref{eq:Eint_penalty_control_new} supplies the required direction ``small penalty residual implies controlled interface contribution,'' but only under the stated PDE-dependent stability assumption.

Conversely, trace estimates explain how local approximation errors create a flux residual. If $f$ is Lipschitz with constant $L_f$ and the exact target jump is $g_{F,i}$, then
\begin{equation}
\begin{aligned}
  |\jump{F(u_h,\partial_xu_h)}_{\Gamma_i}-g_{F,i}|
  &\le L_f(|e_i^-|+|e_{i+1}^+|)\\
  &\quad+\nu(|\partial_{n_i}e_i^-|
  +|\partial_{n_i}e_{i+1}^+|),
\end{aligned}
  \label{eq:flux_factor_pointwise_new}
\end{equation}
where $e_i=u-u_{h,i}$. Equations~\eqref{eq:trace_value} and~\eqref{eq:trace_derivative} then relate the endpoint terms to neighboring $H^1$ and $H^2$ errors. Equations~\eqref{eq:Eint_penalty_control_new} and~\eqref{eq:flux_factor_pointwise_new} state opposite, complementary directions and should not be conflated.

Let $\bm\theta^\star$ minimize the exact-expectation loss within the chosen ansatz and let $\widehat{\bm\theta}$ be the output of a finite-accuracy classical optimizer. Define
\begin{equation}
  \Delta_{\mathrm{opt}}
  =\mathcal L_{\mathrm{tot}}(\widehat{\bm\theta})
  -\mathcal L_{\mathrm{tot}}(\bm\theta^\star)\ge0.
  \label{eq:optimization_gap_new}
\end{equation}
If the squared residual loss is locally stable with constant $C_{\mathrm{opt}}$, then
\begin{equation}
  E_{\mathrm{opt}}
  \le C_{\mathrm{opt}}\sqrt{\Delta_{\mathrm{opt}}}.
  \label{eq:optimization_error_control_new}
\end{equation}
This condition need not hold uniformly in a barren plateau or near a non-identifiable parametrization, which is why optimization conditioning must be measured rather than inferred only from parameter count~\cite{mcclean2018barren,cerezo2021cost,arrasmith2021effect}.

For finite-shot measurements, group the measured residual observables by block $i$ and term $r$. Define the root-mean-square residual-estimation uncertainty
\begin{equation}
  \delta_{\mathrm{shot}}^2
  =\sum_{i,r}\frac{c_{i,r}^2\sigma_{i,r}^2}{M_{i,r}}.
  \label{eq:shot_variance_new}
\end{equation}
Under the same local residual stability,
\begin{equation}
  (\mathbb E E_{\mathrm{shot}}^2)^{1/2}
  \le C_{\mathrm{shot}}\delta_{\mathrm{shot}}.
  \label{eq:shot_error_control_new}
\end{equation}
Thus Eq.~\eqref{eq:shot_variance_new} controls a solution error only after the stability constant has been specified or bounded.

For a fixed partition and a common time discretization, the preceding accounting also gives a transparent sufficient condition at the level of the derived error bounds.  If
\begin{equation}
\begin{aligned}
  &E_x^{\mathrm{blk}}+E_{\mathrm{ans}}^{\mathrm{blk}}
  +E_{\mathrm{opt}}^{\mathrm{blk}}+E_{\mathrm{shot}}^{\mathrm{blk}}
  +\boldsymbol E_{\mathrm{int}}\notag\\
  &\qquad <
  E_x^{\mathrm{uni}}+E_{\mathrm{ans}}^{\mathrm{uni}}
  +E_{\mathrm{opt}}^{\mathrm{uni}}+E_{\mathrm{shot}}^{\mathrm{uni}}.
\end{aligned}
  \label{eq:block_sufficient_comparison}
\end{equation}
then the corresponding block-wise upper bound is smaller than the global upper bound.  Equation~\eqref{eq:block_sufficient_comparison} is a sufficient comparison criterion, not a theorem that the inequality must hold.  The numerical experiments are designed to identify regimes in which the local representation gain exceeds the added interface and optimization costs.

\subsection{Time-adaptive propagation and reblocking}
\label{subsec:time_adaptive_reblocking_new}

For a fixed partition, one time step minimizes Eq.~\eqref{eq:Ltotal_new}:
\begin{equation}
  \{\alpha_i^{m+1},\bm\theta_i^{m+1}\}_{i=1}^K
  =\arg\min\mathcal L_{\mathrm{tot}}^{m+1}.
  \label{eq:one_step_training_new}
\end{equation}
All adaptive indicators are then computed from the reconstructed approximation $u_h(\cdot,t_m)$, not from the exact reference solution.

Let $\mathcal R_{\mathrm{ULA}}^m$ be the union of the current rough/ULA blocks and let $\mathcal R_{\mathrm{rough}}^m$ be obtained from Eq.~\eqref{eq:difficulty_indicator_new}. Define a coverage defect
\begin{equation}
  \eta_{\mathrm{cov}}^m
  =\frac{\int_{\mathcal R_{\mathrm{rough}}^m\setminus
  \mathcal R_{\mathrm{ULA}}^m}\eta(x,t_m)\,\dd x}
  {\int_{\mathcal R_{\mathrm{rough}}^m}\eta(x,t_m)\,\dd x+\varepsilon},
  \label{eq:eta_x_new}
\end{equation}
an ansatz-tail defect
\begin{equation}
  \eta_{\mathrm{ans}}^m
  =\max_i
  \frac{\sum_{k\in\mathcal K_{\mathrm{high}}}|\widehat u_{i,k}^m|^2}
  {\sum_k|\widehat u_{i,k}^m|^2+\varepsilon},
  \label{eq:eta_ans_new}
\end{equation}
and an interface defect
\begin{equation}
  \eta_\Gamma^m
  =\max_{\Gamma_i\in\mathcal G_\Gamma}
  (|\rho_{u,i}|+|\rho_{F,i}|).
  \label{eq:eta_gamma_new}
\end{equation}
A reblocking event occurs when
\begin{equation}
  \eta_{\mathrm{cov}}^m>\theta_{\mathrm{cov}}
  \quad\text{or}\quad
  \eta_{\mathrm{ans}}^m>\theta_{\mathrm{ans}}
  \quad\text{or}\quad
  \eta_\Gamma^m>\theta_\Gamma.
  \label{eq:reblock_trigger_new}
\end{equation}
The Boolean logic is explicitly OR. To prevent chattering, the implementation uses a minimum dwell time $m_{\min}$ between reblocking events and a lower release threshold after a trigger.  The deterministic implementation may additionally cap the displacement of an interface during one event.  For a PDE with known directed transport, a characteristic predictor can prevent a monitor spike from moving the rough block opposite to the propagation direction.  These safeguards use only the PDE coefficients, the old partition, and the current reconstructed state; they do not consult the reference solution.

When triggered, the partition algorithm in Sec.~\ref{subsec:allocation} is applied to $u_h(\cdot,t_m)$, and the old state is transferred by interpolation or $L^2$ projection,
\begin{equation}
\begin{aligned}
  u_{h,\mathrm{new}}^m
  &=\Pi_{\mathrm{old}\to\mathrm{new}}u_{h,\mathrm{old}}^m,\\
  E_{\mathrm{tr}}^m
  &=\|u_{h,\mathrm{old}}^m-u_{h,\mathrm{new}}^m\|_X.
\end{aligned}
  \label{eq:transfer_new}
\end{equation}
The accumulated transfer contribution must be reported separately or absorbed explicitly into the time/ansatz accounting; it is not zero merely because the interface locations are updated.
For an adaptive trajectory with reblocking events $m\in\mathcal E$, a conservative extension of the accounting bound is therefore
\begin{equation}
  E_{\mathrm{tot}}^{\mathrm{adapt}}
  \le E_t+E_x+E_{\mathrm{ans}}+E_{\mathrm{opt}}+E_{\mathrm{shot}}
  +\boldsymbol E_{\mathrm{int}}+\sum_{m\in\mathcal E}E_{\mathrm{tr}}^m,
  \label{eq:adaptive_error_accounting}
\end{equation}
where the sum is deliberately conservative because it ignores possible cancellation between transfer perturbations.

Reallocation at an event must also preserve the purpose of the rough block.  In particular, the production implementation used below does not decrease the rough block's previous collocation count during reblocking; any points needed to enforce this constraint are taken from eligible smooth blocks while respecting the total budget.  Together with the displacement cap, this prevents an apparent adaptive gain from being produced by a collapsing rough interval or by silently removing its local resolution.

\subsection{Quantum circuit realization and loss measurement}
\label{subsec:circuit_measurement_new}

For a smooth block retaining modes $|k|\le M_i$, define the coefficient-register width
\begin{equation}
  m_i=\left\lceil\log_2(2M_i+1)\right\rceil,
  \qquad m_i\le n_i.
  \label{eq:coefficient_register_width_new}
\end{equation}
The coefficient circuit $U_{\mathrm{coef},i}^{(m_i)}$ prepares the centered low-frequency coefficients on $m_i$ qubits. An isometry $V_{\mathrm{emb},i}$ maps the centered frequency labels into the $n_i$-qubit Fourier grid and sets all unretained coefficients to zero. The ZGR-QFT block is therefore
\begin{equation}
  U_i^{\mathrm{ZGR}}
  =\mathrm{QFT}_{n_i}^{\dagger}
  V_{\mathrm{emb},i}
  \left(U_{\mathrm{coef},i}^{(m_i)}\otimes I\right).
  \label{eq:zgr_circuit_factorization}
\end{equation}
This distinguishes the number of coefficient qubits $m_i$, the number of retained modes $M_i$, and the physical-grid width $n_i$.

For a rough block, one concrete layered choice is
\begin{equation}
  U_i^{\mathrm{ULA}}(\bm\theta_i)
  =\prod_{\ell=1}^{d_i}
  \left[
  U_{\mathrm{ent}}
  \prod_{q=1}^{n_i}
  R_z(\theta_{\ell q}^{(3)})
  R_y(\theta_{\ell q}^{(2)})
  R_z(\theta_{\ell q}^{(1)})
  \right],
  \label{eq:ula_circuit}
\end{equation}
where $U_{\mathrm{ent}}$ is a stated nearest-neighbor entangling layer. Circuit order, depth, initialization, and parameter sharing must be held fixed in a resource comparison.

At one time step, let
\begin{equation}
  \ket{r_i(\bm\theta_i)}
  =A_i\ket{u_i(\bm\theta_i)}-\ket{b_i},
  \label{eq:local_residual_state}
\end{equation}
where $A_i$ includes the time-discrete operator and the nonperiodic local derivative closure, and $\ket{b_i}$ contains the known previous-step and source data. Physical boundary conditions remain visible in Eq.~\eqref{eq:Jboundary_new}. Expanding the norm gives
\begin{equation}
\begin{aligned}
  \mathcal L_i
  &=\bra{u_i}A_i^\dagger A_i\ket{u_i}
  -2\operatorname{Re}\bra{b_i}A_i\ket{u_i}
  +\braket{b_i}{b_i}.
\end{aligned}
  \label{eq:local_loss_measurement}
\end{equation}
Pauli decompositions, overlap tests, or Hadamard tests can be used depending on the representation of $A_i$ and $\ket{b_i}$.

Endpoint penalties require signed or complex amplitudes, not merely basis probabilities. From Eq.~\eqref{eq:block_state_new},
\begin{equation}
  u_{i,j}=\alpha_i\braket{j}{\psi_i}.
  \label{eq:endpoint_amplitude_exact}
\end{equation}
A computational-basis sample gives only $|\braket{j}{\psi_i}|^2$. The real and imaginary parts required in Eq.~\eqref{eq:endpoint_amplitude_exact} must be obtained by interference with the known basis state $\ket j$, for example through a Hadamard/overlap test. Near-interface derivatives are then assembled from an order-consistent one-sided stencil; for a second-order right-boundary derivative,
\begin{equation}
  \partial_xu_i(x_i^-)
  \approx\frac{3u_{i,N_i-1}-4u_{i,N_i-2}+u_{i,N_i-3}}{2h_i}.
  \label{eq:boundary_derivative_measurement}
\end{equation}
After endpoint values and derivatives are estimated, the physical flux is assembled classically and inserted into Eqs.~\eqref{eq:normalized_interface_residuals}--\eqref{eq:JGamma_new}. The state-preparation cost, number of Pauli/overlap terms, endpoint-amplitude shots, and repeated block-circuit calls are all part of the total resource cost and must be reported separately from $n_{\mathrm{peak}}$.

\section{Numerical Experiments}
\label{sec:numerics}

This section validates four different components of the proposed block-wise VQA--PDE framework.  Experiment~1 isolates the transmission conditions at a heterogeneous elliptic interface.  Experiment~2 tests fixed block allocation for a linear time-dependent problem and separately examines the expressivity of a concrete ULA circuit.  Experiment~3 addresses a nonlinear PDE and quantifies optimization and finite-shot effects.  Experiment~4 tests whether the monitor, trigger, transfer, and hysteresis rules can follow a moving rough region without access to the reference solution.  In every main PDE experiment, the tested field is obtained by minimizing a discrete PDE residual; the reference field is used only for post-run error evaluation.

The main quantitative conclusions are as follows.  Flux closure reduces the matched-budget error in Experiment~1 from $0.8441$ without flux control to $3.4661\times10^{-3}$.  At the moderate parameter point in Experiment~2, the controlled block method is $33.4\%$ more accurate than the global method, although the global method becomes more accurate at the next tested budget.  In Experiment~3, the controlled block method reduces the error by $76.3\%$, and the finite-shot proxy follows a fitted $M^{-0.4985}$ law.  In Experiment~4, adaptive reblocking improves the time-averaged error over a static partition by $4.70\%$, while hysteresis reduces the event count from 78 to three.  These results delineate the intended operating regime: the block framework is most useful when representation difficulty is localized, whereas a sufficiently large global budget or diffusion-induced smoothing can favor the global ansatz.

\begin{figure}[!htbp]
  \centering
  \includegraphics[width=0.93\textwidth]{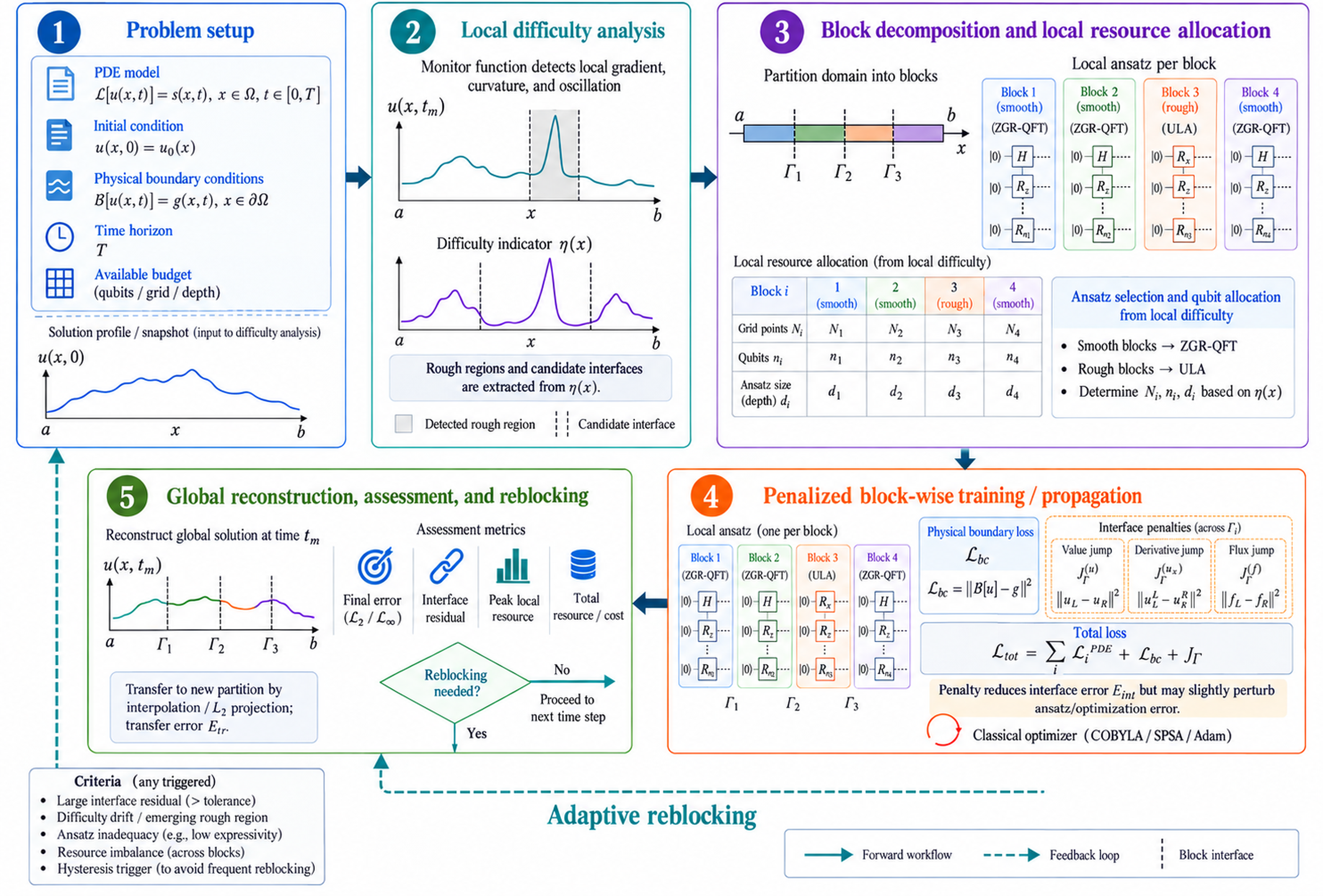}
  \caption{Workflow of the proposed block-wise VQA--PDE framework. Stage~1 defines the PDE, initial and physical-boundary data, time horizon, and available resources. Stage~2 evaluates the normalized local-difficulty monitor and identifies rough regions. Stage~3 places block interfaces and assigns power-of-two grid budgets, qubit widths, and block-dependent ansatz families. Stage~4 minimizes local PDE residuals together with the physical-boundary loss and value, derivative, and PDE-dependent flux interface penalties. Stage~5 reconstructs the global field, evaluates accuracy and resource diagnostics, and triggers adaptive reblocking when the current partition becomes inadequate. A reblocking event transfers the current approximation by interpolation or $L^2$ projection and contributes a transfer error $E_{\mathrm{tr}}$; hysteresis and dwell rules suppress excessive repartitioning. The reference solution is used only for post-run evaluation. The graphical redesign of this schematic used the OpenAI ChatGPT image-generation tool from an author-developed workflow diagram; all scientific content and labels were specified and verified by the author}
  \label{fig:block_workflow}
\end{figure}

\subsection{Protocol, evidence levels, and metrics}
\label{subsec:numerical_protocol}

The experiments separate three evidence levels.  First, the main PDE studies are exact-expectation ansatz-space residual simulations.  Smooth blocks use the lifted low-frequency subspace corresponding to the ZGR--QFT construction, while a rough block uses an expressive local amplitude subspace.  This level tests residual training, representation, allocation, interface closure, and reblocking, but it is not a gate-level simulation.  Second, Experiment~2 independently optimizes the concrete ULA circuit in Eq.~\eqref{eq:ula_circuit} with an exact statevector and an adjoint gradient.  Third, Experiment~3 applies a frozen finite-shot residual-estimator model.  The ULA and shot studies are reported separately and do not turn the main ansatz-space solves into circuit or hardware executions.

The global and block methods use the same total spatial-grid budget.  Parameter comparisons use equal or neighboring total counts, while the largest local grid and parameter counts are reported separately.  Thus a reduction in peak block width is not presented as a reduction in total work.  Physical boundary losses remain active in all no-interface ablations.  Unless stated otherwise, time stepping is backward Euler and artificial interfaces are controlled by both the value jump and the PDE-dependent physical-flux jump.

The primary field metric is
\begin{equation}
  \mathcal E_{L^2}(t)
  =\frac{\|u_{\mathrm{ref}}(\cdot,t)-u_h(\cdot,t)\|_{L^2}}
  {\|u_{\mathrm{ref}}(\cdot,t)\|_{L^2}}.
  \label{eq:numerical_l2_metric}
\end{equation}
We additionally report value and flux jumps, residual evaluations, directly measured optimization and shot deviations, and reblocking transfer errors.  Stochastic studies use ten frozen seeds and report medians with interquartile ranges.  The production runs used Python~3.12 on an Intel Xeon Platinum 8573C CPU.  Source code, frozen settings, seed-level CSV/JSON outputs, tests, and plotting scripts are supplied as Online Resource~1.

\begin{table}[!htbp]
  \caption{Frozen resources for the representative comparisons.  $P$ is the total ansatz-space parameter count, $P_{\max}$ is the largest count in one block, and $n_{\max}$ is the peak local grid width in qubits.  G and B denote global and block solvers; bold B rows mark the proposed block-wise method family.}
  \label{tab:resource_protocol_v8}
  \centering
  \small
  \resizebox{\linewidth}{!}{%
  \begin{tabular}{cccccccc}
    \hline\hline
    Exp. & Method & Grid & Grid per block & $P$ & $P_{\max}$ & $n_{\max}$ & $(\Delta t,T)$\\
    \hline
    1 & G & 128 & 128 & 8 & 8 & 7 & steady\\
    \bfseries 1 & \bfseries B & \bfseries 128 & \bfseries $64+64$ & \bfseries 8 & \bfseries 4 & \bfseries 6 & \bfseries steady\\
    2 & G & 128 & 128 & 43 & 43 & 7 & $(5\!\times\!10^{-4},0.06)$\\
    \bfseries 2 & \bfseries B & \bfseries 128 & \bfseries $32+64+32$ & \bfseries 44 & \bfseries 28 & \bfseries 6 & \bfseries $(5\!\times\!10^{-4},0.06)$\\
    3 & G & 64 & 64 & 33 & 33 & 6 & $(10^{-3},0.02)$\\
    \bfseries 3 & \bfseries B & \bfseries 64 & \bfseries $16+32+16$ & \bfseries 34 & \bfseries 22 & \bfseries 5 & \bfseries $(10^{-3},0.02)$\\
    4 & G & 128 & 128 & 43 & 43 & 7 & $(10^{-3},0.35)$\\
    \bfseries 4 & \bfseries B & \bfseries 128 & \bfseries $32+64+32$ & \bfseries 44 & \bfseries 28 & \bfseries 6 & \bfseries $(10^{-3},0.35)$\\
    \hline\hline
  \end{tabular}%
  }
\end{table}

\subsection{Experiment 1: heterogeneous elliptic transmission problem}
\label{subsec:exp1_v8}

\noindent\textbf{Setup.}\ We solve
\begin{equation}
  -\partial_x\!\left(\kappa(x)u_x\right)=f(x),\qquad
  u(0)=u(1)=0,
  \label{eq:exp1_pde_v8}
\end{equation}
where $\kappa=1$ on $[0,1/2]$ and $\kappa=1/4$ on $(1/2,1]$.  The manufactured solution and forcing are
\begin{align}
 u_{\mathrm{ex}}(x)&=
 \begin{cases}
 \sin(4\pi x)-1.6\pi x,&x\le 1/2,\\
 2\sin(4\pi x)+1.6\pi(x-1),&x>1/2,
 \end{cases}
 \label{eq:exp1_exact_v8}\\
 f(x)&=
 \begin{cases}
 16\pi^2\sin(4\pi x),&x\le 1/2,\\
 8\pi^2\sin(4\pi x),&x>1/2.
 \end{cases}
\end{align}
Both $u$ and the physical flux $F=-\kappa u_x$ are continuous at $x=1/2$, so the coefficient jump does not conceal a delta source.  The block interface is placed at the known coefficient discontinuity, independently of $u_{\mathrm{ex}}$.  A conservative second-order flux discretization is used globally, while block traces use second-order one-sided closures.

\noindent\textbf{Results and interpretation.}\ At the matched budget $P=8$, the global error is $1.7554$, the block error without flux control is $0.8441$, and the value-plus-flux block error is $3.4661\times10^{-3}$.  A value-only penalty drives the value jump to roundoff but leaves a flux jump of $6.3092$; adding the physical-flux condition reduces the value and flux jumps to $5.55\times10^{-14}$ and $4.79\times10^{-13}$.  With $P=120$, the global error decreases to $8.9591\times10^{-3}$, showing that the global discretization is consistent but requires a much larger representation budget for this interface profile.

\begin{figure}[!htbp]
  \centering
  \makebox[\textwidth][c]{\includegraphics[height=\ResultPanelHeight,keepaspectratio]{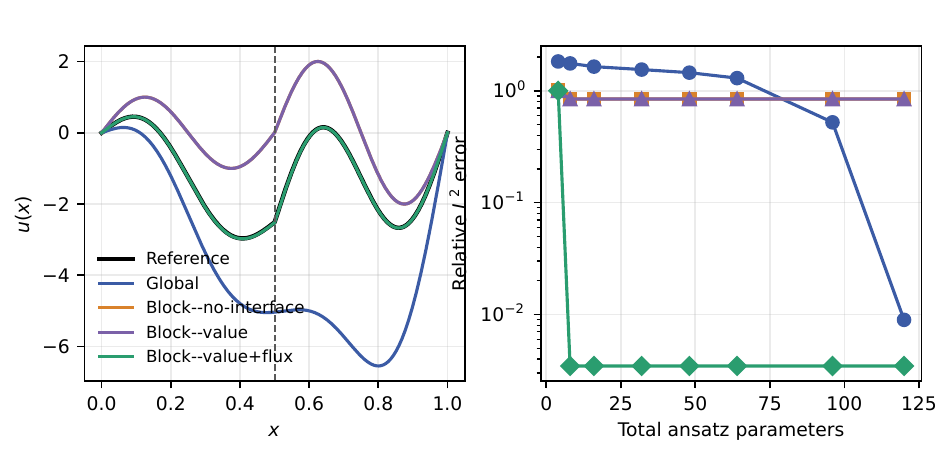}}

  \vspace{0.55em}
  \makebox[\textwidth][c]{%
    \includegraphics[height=\ResultPanelHeight,keepaspectratio]{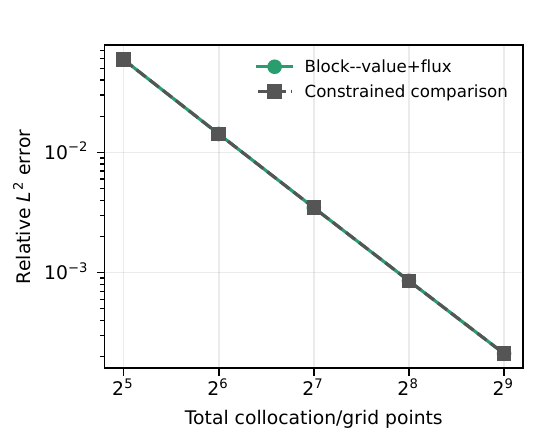}%
    \hspace{0.8em}%
    \includegraphics[height=\ResultPanelHeight,keepaspectratio]{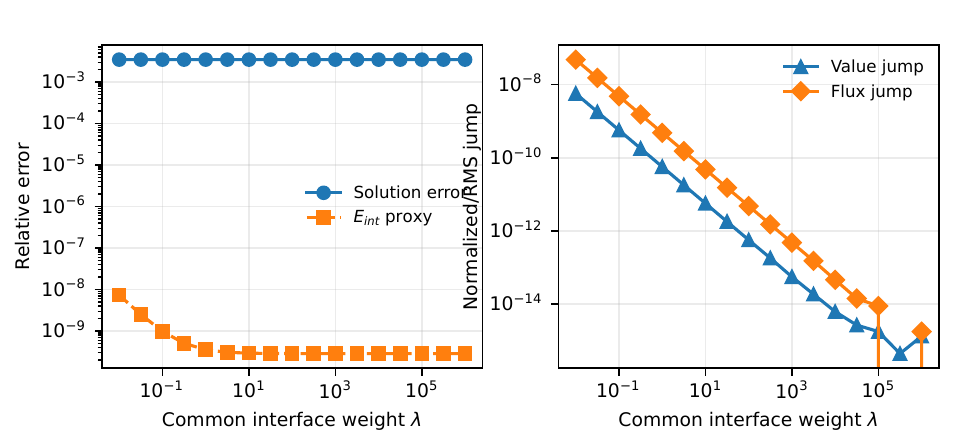}%
  }
  \caption{Experiment~1, shown at the fixed panel height used throughout this section.  Top left: reference, global, and flux-controlled block profiles at the matched budget; the dashed line is the material interface.  Top right: relative error versus total ansatz parameters.  Bottom left: closure/grid convergence, with approximately second-order decay.  Bottom center: field error versus the common interface weight.  Bottom right: RMS value and physical-flux jumps versus that weight.  The unstable no-flux profile is omitted from the profile scale but retained in the numerical comparison}
  \label{fig:exp1_results_v8}
\end{figure}

Figure~\ref{fig:exp1_results_v8} presents the Experiment~1 evidence in the same order as the argument above.  The profile panel shows that the controlled block field follows the exact piecewise-smooth solution across $x=1/2$, and the parameter curve separates the low-budget block advantage from the eventual improvement of the global basis.  The grid curve verifies that the controlled result is not merely a single-grid fit: the errors at 32, 64, 128, 256, and 512 total points decrease by approximately a factor of four per refinement.  The interface-residual panel shows that increasing the common penalty weight drives both value and flux jumps down by many orders of magnitude.  The corresponding field error remains near the spatial-discretization floor over the tested range.

\subsection{Experiment 2: linear advection--diffusion}
\label{subsec:exp2_v8}

\begin{figure}[!htbp]
  \centering
  \makebox[\textwidth][c]{\includegraphics[height=\ResultPanelHeight,keepaspectratio]{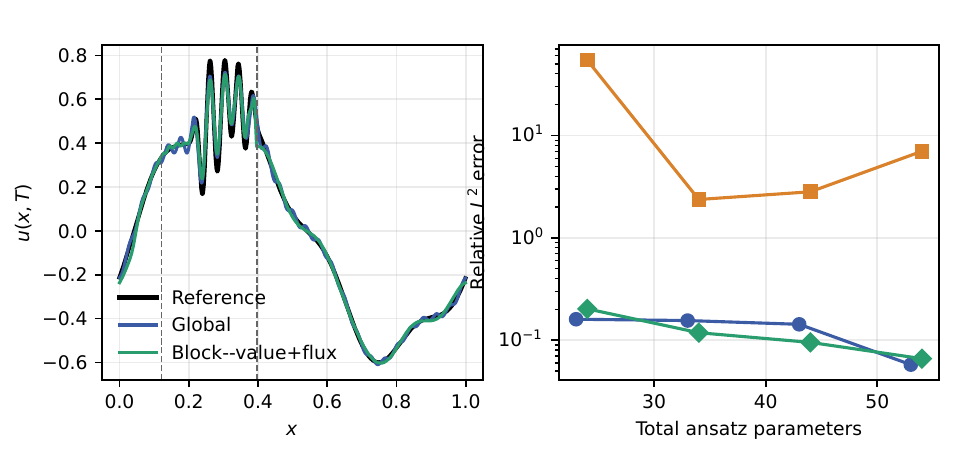}}

  \vspace{0.45em}
  \makebox[\textwidth][c]{\includegraphics[height=\ResultPanelHeight,keepaspectratio]{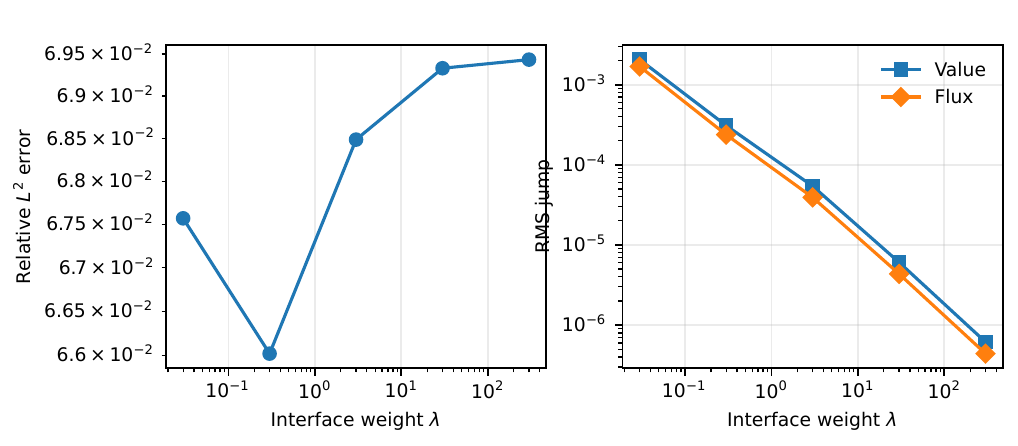}}

  \vspace{0.45em}
  \makebox[\textwidth][c]{\includegraphics[height=\ResultPanelHeight,keepaspectratio]{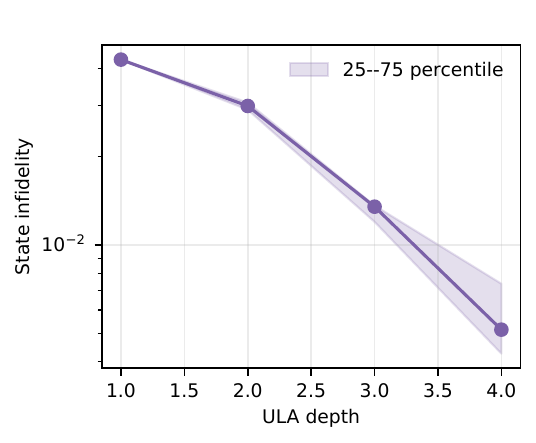}}
  \caption{Experiment~2, with every ordinary panel shown at the common fixed height.  First row, left: final reference, global, and controlled block profiles; dashed lines delimit the rough block selected from $u_0$.  First row, right: relative error versus total ansatz parameters, including the moderate-budget block gain and higher-budget global crossover.  Second row: field error and interface jumps versus the common value/flux weight.  Third row: the separate five-qubit ULA statevector check; points are medians over ten initializations and the band is the 25--75 percentile range}
  \label{fig:exp2_residual_v8}
  \label{fig:ula_depth_v8}
\end{figure}

\noindent\textbf{Setup.}\ On the periodic unit interval we solve
\begin{equation}
  u_t+0.7u_x=2\times10^{-4}u_{xx},
  \label{eq:exp2_pde_v8}
\end{equation}
with
\begin{align}
 u_0(x)={}&0.55\sin(2\pi x)+0.08\sin(8\pi x)
 \notag\\
 &+0.30w(x)\sin(48\pi x)+0.08w(x)\cos(36\pi x),
 \label{eq:exp2_ic_v8}\\
 w(x)={}&\tfrac12\!\left[\tanh\!\frac{x-0.18}{0.012}
 -\tanh\!\frac{x-0.34}{0.012}\right].
\end{align}
The monitor evaluated only on $u_0$ gives the partition $[0,0.121015625,0.3965625,1]$, the allocation $[32,64,32]$, and the family sequence ZGR--rough--ZGR.  The reference is the exact Fourier propagator, whereas the tested solvers minimize the backward-Euler residual with $\Delta t=5\times10^{-4}$.  The backward-Euler discretization error relative to the Fourier reference is $2.4413\times10^{-2}$ and therefore forms a visible floor in the ansatz comparison.

\noindent\textbf{Residual-solver results.}\ At neighboring total counts $P=43$ globally and $P=44$ block-wise, the errors are $0.14291$ and $0.09517$; the block method is $33.4\%$ lower and uses six rather than seven peak grid qubits.  Its RMS value and physical-flux jumps are $8.36\times10^{-4}$ and $6.17\times10^{-4}$.  The no-interface field is unstable, with error $2.8173$.  At the next tested counts, $P=53$ and $P=54$, the global error $0.05771$ is smaller than the block error $0.06602$.  The result therefore supports a moderate-budget gain rather than universal dominance.

In Fig.~\ref{fig:exp2_residual_v8}, the upper-left panel locates the rough packet between the two dashed interfaces and shows that the global and controlled block fields are both stable.  The upper-right parameter curve is the key fairness check: the controlled block point is better near $P=44$, while the global curve crosses below it near $P=53$.  The lower-left penalty panel shows a broad accuracy minimum around the frozen choice $\lambda=0.3$.  The lower-right panel shows that larger weights continue to reduce the interface jumps, but with a slight field-error increase.  This is the numerical manifestation of the tradeoff described after Eq.~\eqref{eq:Ltotal_new}: the penalty primarily suppresses the interface contribution, while the changed constrained optimum can slightly perturb the realized ansatz/optimization error.  Interface consistency and field accuracy are therefore related but not identical objectives.

\noindent\textbf{Concrete ULA statevector check.}\ The separate five-qubit ULA study uses ten initializations at each depth and a fixed optimization budget.  Median infidelities at depths one through four are $4.289\times10^{-2}$, $2.980\times10^{-2}$, $1.351\times10^{-2}$, and $5.143\times10^{-3}$.  Figure~\ref{fig:ula_depth_v8} shows both the monotone median reduction and the seed-to-seed interquartile band.  This supports increasing expressivity for the target rough-block state under the tested budget; it is not a proof of global optimization or a gate-level simulation of the full time trajectory.

\subsection{Experiment 3: viscous Burgers equation}
\label{subsec:exp3_v8}

\begin{figure}[!htbp]
  \centering
  \makebox[\textwidth][c]{\includegraphics[height=\ResultPanelHeight,keepaspectratio]{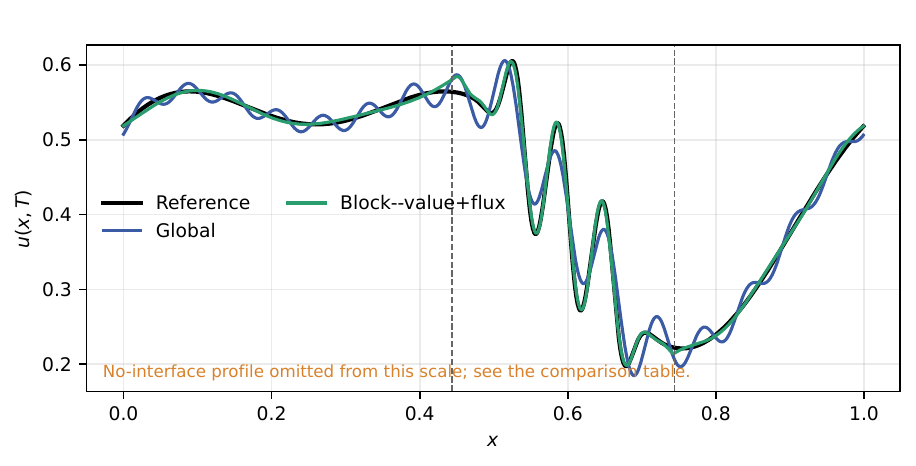}}

  \vspace{0.55em}
  \makebox[\textwidth][c]{\includegraphics[height=\ResultPanelHeight,keepaspectratio]{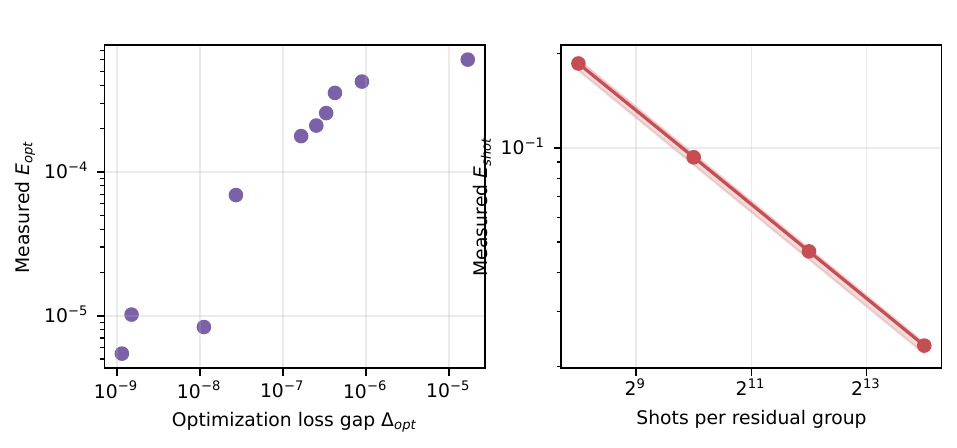}}
  \caption{Experiment~3, with both rows shown at the common fixed height.  Top: nonlinear solution comparison at $T=0.02$; the black, blue, and green curves are the reference, global, and value-plus-flux block solutions, and dashed lines mark the rough-block interfaces.  Bottom left: directly measured optimizer deviation $E_{\mathrm{opt}}$ versus loss gap $\Delta_{\mathrm{opt}}$ for ten fixed-budget seeds.  Bottom right: median shot-induced deviation versus shots per residual group; faint lines show individual seeds and the central trend follows $M^{-1/2}$.  The divergent no-interface profile is omitted from the plotted range but retained in the error table.}
  \label{fig:exp3_solution_v8}
  \label{fig:exp3_noise_v8}
\end{figure}

\noindent\textbf{Setup.}\ The nonlinear test is
\begin{equation}
\begin{aligned}
  u_t+\partial_x(u^2/2)&=0.003u_{xx},\\
  F(u,u_x)&=u^2/2-0.003u_x,
\end{aligned}
  \label{eq:exp3_pde_v8}
\end{equation}
with periodic boundaries and
\begin{align}
u_0(x)={}&0.45+0.15\sin(2\pi x)+0.08\cos(4\pi x)
\notag\\
&+0.18w_B(x)\sin(32\pi x),
\end{align}
where $w_B$ has the same form as Eq.~\eqref{eq:exp2_ic_v8} but endpoints $0.50$ and $0.68$.  The initial monitor produces the partition $[0,0.44328125,0.74421875,1]$ and allocation $[16,32,16]$.  The reference is a 1024-point, two-thirds-dealiased Fourier pseudospectral solution advanced by fourth-order Runge--Kutta with internal step $10^{-5}$.  The tested solver uses $\Delta t=10^{-3}$ and nonlinear backward-Euler residual minimization.

\noindent\textbf{Matched-resource nonlinear solve.}\ With neighboring parameter counts 33 and 34, the global and flux-controlled block errors are $3.8112\times10^{-2}$ and $9.0360\times10^{-3}$, respectively, a $76.3\%$ reduction.  Peak grid width falls from six to five qubits.  The controlled RMS value and nonlinear-flux jumps are $8.16\times10^{-6}$ and $2.31\times10^{-6}$.  Removing the interface terms makes the local subproblems underconstrained and unstable: the field error is $7.2031$, with value and flux jumps $40.2$ and $1.45\times10^3$.

Figure~\ref{fig:exp3_solution_v8} shows where the gain occurs.  The controlled block field follows the reference through the oscillatory middle interval, while the global field exhibits visible phase and amplitude deviations on both smooth and rough parts of the domain.  The two dashed lines are the monitor-selected interfaces.  The unstable no-interface profile lies far outside this plotting range and is intentionally represented by its numerical error rather than by compressing the useful curves.

\noindent\textbf{Optimization and finite-shot deviations.}\ To isolate $E_{\mathrm{opt}}$, ten seeded runs solve the same fixed final backward-Euler residual problem with only three residual evaluations and are compared with a 120-evaluation solution.  The median directly measured $E_{\mathrm{opt}}$ is $1.935\times10^{-4}$ with interquartile range $[2.489\times10^{-5},3.294\times10^{-4}]$; the median loss gap is $2.090\times10^{-7}$.  These deliberately truncated optimizations are used only to expose optimizer variability.

For the shot study, every residual group receives frozen, independent zero-mean Gaussian estimator noise with standard deviation $0.10/\sqrt{M}$.  Ten full noisy trajectories at $M=256,1024,4096,$ and $16384$ give median $E_{\mathrm{shot}}$ values $0.18576$, $0.09323$, $0.04670$, and $0.02337$.  The right panel of Fig.~\ref{fig:exp3_noise_v8} is nearly linear on logarithmic axes and has fitted slope $-0.4985$, consistent with the prescribed $M^{-1/2}$ estimator scaling.  This is a measurement-statistics proxy, not a gate-noise or hardware-noise model.

\subsection{Experiment 4: adaptive reblocking}
\label{subsec:exp4_v8}

\begin{figure}[!htbp]
  \centering
  \makebox[\textwidth][c]{\includegraphics[height=\ResultPanelHeight,keepaspectratio]{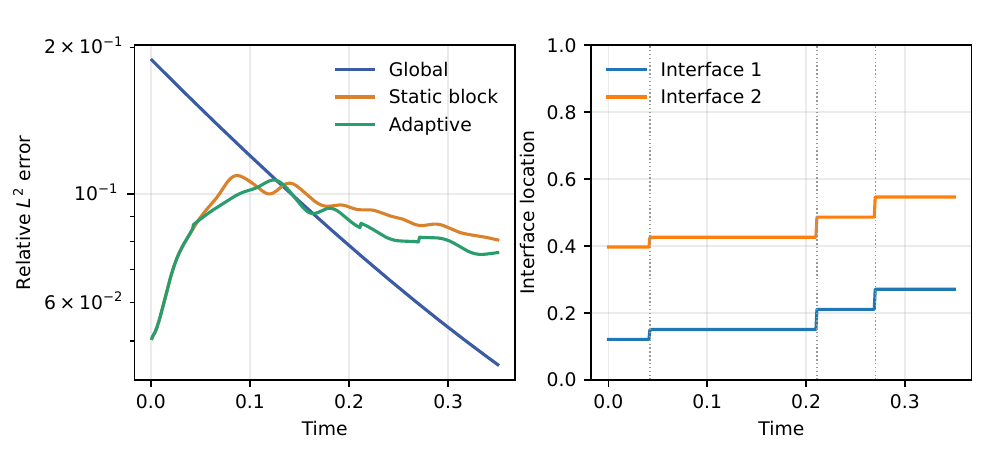}}

  \vspace{0.45em}
  \makebox[\textwidth][c]{\includegraphics[height=0.36\textheight,keepaspectratio]{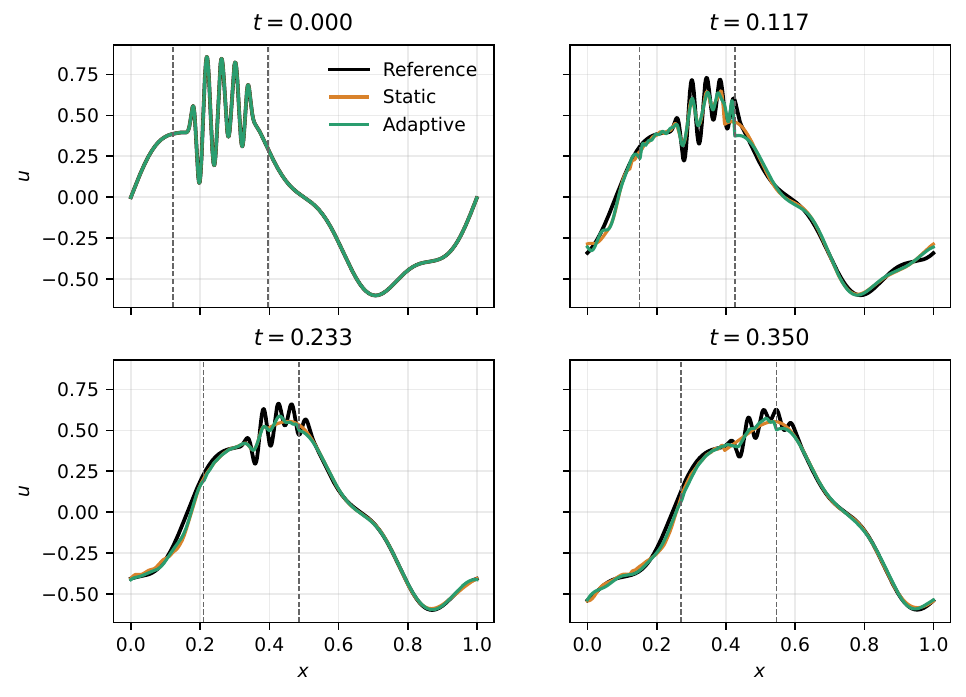}}

  \vspace{0.45em}
  \makebox[\textwidth][c]{\includegraphics[height=\ResultPanelHeight,keepaspectratio]{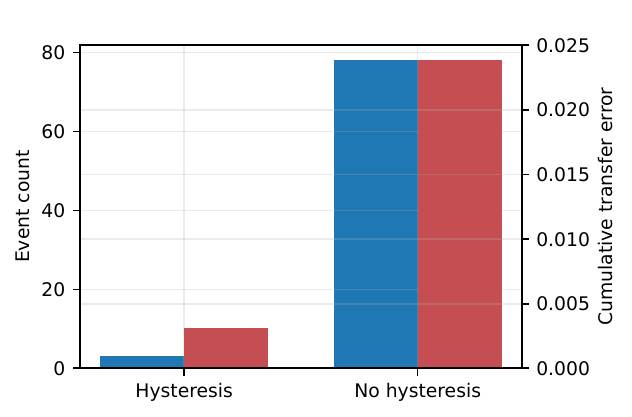}}
  \caption{Experiment~4 results grouped with their governing equations and discussion.  Top: relative errors and stepwise interface trajectories; dotted vertical lines mark the three reblocking events.  Middle: four field snapshots, with dashed adaptive interfaces.  Bottom: event count and cumulative transfer error with and without hysteresis.  Panel heights follow the common convention used in the preceding experiments}
  \label{fig:exp4_history_v8}
  \label{fig:exp4_snapshots_v8}
  \label{fig:exp4_hysteresis_v8}
\end{figure}

\noindent\textbf{Setup.}\ The final experiment reuses Eqs.~\eqref{eq:exp2_pde_v8}--\eqref{eq:exp2_ic_v8} but extends the horizon to $T=0.35$ with $\Delta t=10^{-3}$.  Global, static-block, adaptive-block, and no-hysteresis adaptive solvers share the 128-point budget.  The adaptive indicators are computed exclusively from the current reconstructed field.  The frozen thresholds are $\theta_{\mathrm{cov}}=0.15$, $\theta_{\mathrm{ans}}=0.05$, and $\theta_\Gamma=0.015$, combined by the OR rule in Eq.~\eqref{eq:reblock_trigger_new}; the dwell time is 12 steps and the release factor is $0.55$.

The implementation uses cubic interpolation for transfer, caps each rough-block center displacement at $0.06$, applies the known positive-advection characteristic as a directional safeguard, and does not reduce the rough block's 64-point resolution during reallocation.  These deterministic constraints use no reference data and prevent a monitor spike from collapsing or teleporting the rough block.

\noindent\textbf{Accuracy and interface motion.}\ The time-averaged errors are $0.09618$ globally, $0.09037$ for the static block method, $0.08612$ for adaptive reblocking, and $0.08833$ without hysteresis.  The hysteretic adaptive method therefore improves the static method by $4.70\%$.  It reblocks at $t=0.042$, $0.211$, and $0.270$, with cumulative relative transfer error $3.141\times10^{-3}$.

The left panel of Fig.~\ref{fig:exp4_history_v8} shows that the adaptive curve is below the static-block curve over most of the later trajectory, which is the basis for the time-averaged improvement.  The global curve eventually becomes the smallest because diffusion progressively removes the localized high-frequency structure.  The right panel connects each discontinuous interface displacement to one of the three event times; between events, the partition remains fixed, so the method is reblocking rather than continuously moving the mesh.

\noindent\textbf{Field snapshots and endpoint behavior.}\ At final time, the global, static, adaptive, and no-hysteresis errors are $0.04474$, $0.08058$, $0.07591$, and $0.07798$.  Figure~\ref{fig:exp4_snapshots_v8} explains why the time-average and endpoint conclusions differ.  At early and intermediate times, the adaptive interfaces move with the oscillatory packet and the adaptive reconstruction is slightly closer to the reference than the static reconstruction.  By $t=0.35$, diffusion has smoothed the packet and a global Fourier representation is again favorable.  The adaptive claim is therefore based on the full trajectory and controlled event history, not on final-time superiority.

\noindent\textbf{Hysteresis ablation.}\ Removing hysteresis produces 78 events instead of three, increases cumulative transfer error from $3.141\times10^{-3}$ to $2.383\times10^{-2}$, and slightly worsens the time-averaged field error.  Figure~\ref{fig:exp4_hysteresis_v8} makes the two consequences visible on separate axes: the blue bars measure event count and the red bars measure accumulated transfer error.  The result supports the release threshold and minimum dwell time as numerical safeguards, not merely implementation details.

\begin{table}[!htbp]
  \caption{Experiment~4 trajectory summary. Mean and final errors use the relative $L^2$ metric in Eq.~\eqref{eq:numerical_l2_metric}; transfer error is accumulated only when reblocking occurs. Bold rows are block-wise variants of the proposed method family}
  \label{tab:exp4_summary_latest}
  \centering
  \small
\resizebox{\linewidth}{!}{%
  \begin{tabular}{lcccc}
    \hline\hline
    Method & Mean error & Final error & Events & Cumulative transfer \\
    \hline
    Global & $0.09618$ & $0.04474$ & 0 & 0 \\
    \textbf{Static block} & \textbf{$0.09037$} & \textbf{$0.08058$} & \textbf{0} & \textbf{0} \\
    \textbf{Adaptive block} & \textbf{$0.08612$} & \textbf{$0.07591$} & \textbf{3} & \textbf{$3.141\times10^{-3}$} \\
    \textbf{Adaptive, no hysteresis} & \textbf{$0.08833$} & \textbf{$0.07798$} & \textbf{78} & \textbf{$2.383\times10^{-2}$} \\
    \hline\hline
  \end{tabular}%
  }
\end{table}

\subsection{Cross-experiment interpretation and limitations}
\label{subsec:numerical_discussion_v8}

\begin{table}[!htbp]
  \caption{Frozen comparison points; Experiment~4 reports time averages. Both the controlled-block and no-interface columns are variants of the proposed block-wise method and are bolded; the latter is an interface-ablation diagnostic}
  \label{tab:main_results_v8}
  \centering
  \small
\resizebox{\linewidth}{!}{%
  \begin{tabular}{clcccc}
    \hline\hline
    Exp. & Comparison & Global & \textbf{Controlled block} & \textbf{No interface} & Interpretation\\
    \hline
    1 & $\mathcal E_{L^2}$, $P=8$ & $1.7554$ & \textbf{$3.4661\times10^{-3}$} & \textbf{$0.8441$} & flux closure essential\\
    2 & $\mathcal E_{L^2}$, $P=43/44$ & $0.14291$ & \textbf{$0.09517$} & \textbf{$2.8173$} & moderate-budget gain\\
    3 & $\mathcal E_{L^2}$, $P=33/34$ & $0.03811$ & \textbf{$0.009036$} & \textbf{$7.2031$} & nonlinear gain\\
    4 & mean $\mathcal E_{L^2}$ & $0.09618$ & \textbf{$0.08612$} & \textbf{static: $0.09037$} & modest adaptive gain\\
    \hline\hline
  \end{tabular}%
  }
\end{table}

Taken together, the experiments support the mechanisms claimed in Sec.~\ref{sec:block_method_error}: local representation can help when difficult structure is spatially concentrated; physical-flux penalties are necessary to close independently trained blocks; finite optimizer and measurement budgets create distinct measurable deviations; and hysteretic reblocking can follow a moving rough region without reference access.  They also identify two limits.  The global ansatz overtakes the block method at the higher Experiment~2 parameter point and at the smoothed final state of Experiment~4.  Moreover, the main residual solver represents the rough block in an expressive amplitude subspace, while only a separate target-state test uses the concrete ULA circuit.

Consequently, these results establish neither quantum advantage nor universal block superiority.  A gate-level study must still compile every residual observable, optimize the full circuit trajectory under finite shots, include device noise, and compare total execution cost.  The present evidence is a reproducible validation of the proposed decomposition, interface, error-accounting, and adaptation logic at explicitly stated simulation levels.

\section{Conclusion and Outlook}
\label{sec:conclusion}
In this work, we proposed a block-wise variational quantum algorithm (VQA) framework for solving partial differential equations (PDEs) with spatially nonuniform solution complexity.

Methodologically, the framework integrates a computable local monitor and deterministic power-of-two resource allocation, adopting block-dependent ansatz families.
By rigorously separating physical-boundary and artificial-interface residuals, and incorporating PDE-dependent flux closure with explicit transfer errors via hysteretic reblocking, we systematically disentangle spatial, ansatz, optimization, shot, interface, time-discretization, and transfer contributions.
This approach prevents the erroneous inference of solution bounds from small training losses alone and establishes a rigorous theoretical foundation for residual-to-solution control, invoked exclusively under a stated PDE-dependent stability assumption.

Numerically, four reproducible residual-trained experiments demonstrate the conditional utility of the proposed method.
The heterogeneous elliptic test confirms that value continuity alone is insufficient for convergence and that physical-flux control is essential to restore the correct transmission solution.
Under matched or neighboring budgets, the controlled block method effectively reduces reported errors in the moderate-budget linear test and achieves a 76.3\% reduction for the nonlinear Burgers test, while simultaneously decreasing the peak grid width by one qubit.
However, at larger linear budgets, the global method outperforms the block-wise approach, clarifying that the proposed method is not universally superior.
Furthermore, a concrete five-qubit uniform linear array (ULA) target-state study shows performance improvement with increasing circuit depth, and the finite-shot residual proxy is shown to follow a fitted $M^{-0.4985}$ scaling law.
The adaptive test yields a 4.70\% improvement in time-averaged error over static partitioning and successfully reduces uncontrolled events from 78 to 3 via hysteresis.
Nevertheless, under diffusive smoothing, the global method performs best at the final time, ruling out any universal block-superiority interpretation.

For future work and in conclusion, before strong hardware-oriented claims can be made, a unified circuit-level implementation is required.
This implementation must optimize the same compiled local residuals, endpoint observables, flux penalties, and time trajectories under finite shots and realistic device noise.
Future studies should comprehensively report total circuit executions, circuit depths, classical optimization costs, communication and reconstruction overhead, and peak widths.
Moreover, extensions to higher dimensions will necessitate geometrically robust partitions and interface quadrature.
Within these explicitly stated limitations and conditional conclusions, the methodology presented herein provides a reproducible and robust baseline for testing useful locality in VQA-based PDE solvers.

\section*{Acknowledgments}

This work is supported by the Fundamental Research Funds for the Central Universities (Grants No. 3072025YC2404).


\begin{thebibliography}{49}
\providecommand{\natexlab}[1]{#1}
\providecommand{\url}[1]{\texttt{#1}}
\expandafter\ifx\csname urlstyle\endcsname\relax
  \providecommand{\doi}[1]{doi: #1}\else
  \providecommand{\doi}{doi: \begingroup \urlstyle{rm}\Url}\fi

\bibitem[Peruzzo et~al.(2014)Peruzzo, McClean, Shadbolt, Yung, Zhou, Love,
  Aspuru-Guzik, and O'Brien]{peruzzo2014variational}
Alberto Peruzzo, Jarrod McClean, Peter Shadbolt, Man-Hong Yung, Xiao-Qi Zhou,
  Peter~J. Love, Al{\'a}n Aspuru-Guzik, and Jeremy~L. O'Brien.
\newblock A variational eigenvalue solver on a photonic quantum processor.
\newblock \emph{Nature Communications}, 5:\penalty0 4213, 2014.
\newblock \doi{10.1038/ncomms5213}.

\bibitem[Farhi et~al.(2014)Farhi, Goldstone, and Gutmann]{farhi2014qaoa}
Edward Farhi, Jeffrey Goldstone, and Sam Gutmann.
\newblock A quantum approximate optimization algorithm.
\newblock \emph{arXiv preprint arXiv:1411.4028}, 2014.

\bibitem[Preskill(2018)]{preskill2018nisq}
John Preskill.
\newblock Quantum computing in the {NISQ} era and beyond.
\newblock \emph{Quantum}, 2:\penalty0 79, 2018.
\newblock \doi{10.22331/q-2018-08-06-79}.

\bibitem[Cerezo et~al.(2021{\natexlab{a}})Cerezo, Arrasmith, Babbush, Benjamin,
  Endo, Fujii, McClean, Mitarai, Yuan, Cincio, and
  Coles]{cerezo2021variational}
M.~Cerezo, Andrew Arrasmith, Ryan Babbush, Simon~C. Benjamin, Suguru Endo,
  Keisuke Fujii, Jarrod~R. McClean, Kosuke Mitarai, Xiao Yuan, Lukasz Cincio,
  and Patrick~J. Coles.
\newblock Variational quantum algorithms.
\newblock \emph{Nature Reviews Physics}, 3:\penalty0 625--644,
  2021{\natexlab{a}}.
\newblock \doi{10.1038/s42254-021-00348-9}.

\bibitem[Bharti et~al.(2022)Bharti, Cervera-Lierta, Kyaw, Haug, Alperin-Lea,
  Anand, Degroote, Heimonen, Kottmann, Menke, Mok, Sim, Kwek, and
  Aspuru-Guzik]{bharti2022noisy}
Kishor Bharti, Alba Cervera-Lierta, Thi~Ha Kyaw, Tobias Haug, Sumner
  Alperin-Lea, Abhinav Anand, Matthias Degroote, Hermanni Heimonen, Jakob~S.
  Kottmann, Tim Menke, Wai-Keong Mok, Sukin Sim, Leong-Chuan Kwek, and Al{\'a}n
  Aspuru-Guzik.
\newblock Noisy intermediate-scale quantum algorithms.
\newblock \emph{Reviews of Modern Physics}, 94:\penalty0 015004, 2022.
\newblock \doi{10.1103/RevModPhys.94.015004}.

\bibitem[Tilly et~al.(2022)Tilly, Chen, Cao, Picozzi, Setia, Li, Grant,
  Wossnig, Rungger, Booth, and Tennyson]{tilly2022variational}
Jules Tilly, Hongxiang Chen, Shuxiang Cao, Dario Picozzi, Kanav Setia, Ying Li,
  Edward Grant, Leonard Wossnig, Ivan Rungger, George~H. Booth, and Jonathan
  Tennyson.
\newblock The variational quantum eigensolver: A review of methods and best
  practices.
\newblock \emph{Physics Reports}, 986:\penalty0 1--128, 2022.
\newblock \doi{10.1016/j.physrep.2022.08.003}.

\bibitem[McClean et~al.(2018)McClean, Boixo, Smelyanskiy, Babbush, and
  Neven]{mcclean2018barren}
Jarrod~R. McClean, Sergio Boixo, Vadim~N. Smelyanskiy, Ryan Babbush, and
  Hartmut Neven.
\newblock Barren plateaus in quantum neural network training landscapes.
\newblock \emph{Nature Communications}, 9:\penalty0 4812, 2018.
\newblock \doi{10.1038/s41467-018-07090-4}.

\bibitem[Cerezo et~al.(2021{\natexlab{b}})Cerezo, Sone, Volkoff, Cincio, and
  Coles]{cerezo2021cost}
M.~Cerezo, Akira Sone, Tyler Volkoff, Lukasz Cincio, and Patrick~J. Coles.
\newblock Cost function dependent barren plateaus in shallow parametrized
  quantum circuits.
\newblock \emph{Nature Communications}, 12:\penalty0 1791, 2021{\natexlab{b}}.
\newblock \doi{10.1038/s41467-021-21728-w}.

\bibitem[Arrasmith et~al.(2021)Arrasmith, Cerezo, Czarnik, Cincio, and
  Coles]{arrasmith2021effect}
Andrew Arrasmith, M.~Cerezo, Piotr Czarnik, Lukasz Cincio, and Patrick~J.
  Coles.
\newblock Effect of barren plateaus on gradient-free optimization.
\newblock \emph{Quantum}, 5:\penalty0 558, 2021.
\newblock \doi{10.22331/q-2021-10-05-558}.

\bibitem[Grant et~al.(2019)Grant, Wossnig, Ostaszewski, and
  Benedetti]{grant2019initialization}
Edward Grant, Leonard Wossnig, Mateusz Ostaszewski, and Marcello Benedetti.
\newblock An initialization strategy for addressing barren plateaus in
  parametrized quantum circuits.
\newblock \emph{Quantum}, 3:\penalty0 214, 2019.
\newblock \doi{10.22331/q-2019-12-09-214}.

\bibitem[Schuld et~al.(2019)Schuld, Bergholm, Gogolin, Izaac, and
  Killoran]{schuld2019evaluating}
Maria Schuld, Ville Bergholm, Christian Gogolin, Josh Izaac, and Nathan
  Killoran.
\newblock Evaluating analytic gradients on quantum hardware.
\newblock \emph{Physical Review A}, 99:\penalty0 032331, 2019.
\newblock \doi{10.1103/PhysRevA.99.032331}.

\bibitem[Mitarai et~al.(2018)Mitarai, Negoro, Kitagawa, and
  Fujii]{mitarai2018quantum}
Kosuke Mitarai, Makoto Negoro, Masahiro Kitagawa, and Keisuke Fujii.
\newblock Quantum circuit learning.
\newblock \emph{Physical Review A}, 98:\penalty0 032309, 2018.
\newblock \doi{10.1103/PhysRevA.98.032309}.

\bibitem[Huang et~al.(2020)Huang, Kueng, and Preskill]{huang2020predicting}
Hsin-Yuan Huang, Richard Kueng, and John Preskill.
\newblock Predicting many properties of a quantum system from very few
  measurements.
\newblock \emph{Nature Physics}, 16:\penalty0 1050--1057, 2020.
\newblock \doi{10.1038/s41567-020-0932-7}.

\bibitem[Harrow et~al.(2009)Harrow, Hassidim, and Lloyd]{harrow2009quantum}
Aram~W. Harrow, Avinatan Hassidim, and Seth Lloyd.
\newblock Quantum algorithm for linear systems of equations.
\newblock \emph{Physical Review Letters}, 103:\penalty0 150502, 2009.
\newblock \doi{10.1103/PhysRevLett.103.150502}.

\bibitem[Childs et~al.(2017)Childs, Kothari, and Somma]{childs2017quantum}
Andrew~M. Childs, Robin Kothari, and Rolando~D. Somma.
\newblock Quantum algorithm for systems of linear equations with exponentially
  improved dependence on precision.
\newblock \emph{SIAM Journal on Computing}, 46\penalty0 (6):\penalty0
  1920--1950, 2017.
\newblock \doi{10.1137/16M1087072}.

\bibitem[Berry(2014)]{berry2014high}
Dominic~W. Berry.
\newblock High-order quantum algorithm for solving linear differential
  equations.
\newblock \emph{Journal of Physics A: Mathematical and Theoretical},
  47:\penalty0 105301, 2014.
\newblock \doi{10.1088/1751-8113/47/10/105301}.

\bibitem[Clader et~al.(2013)Clader, Jacobs, and
  Sprouse]{clader2013preconditioned}
B.~David Clader, Bryan~C. Jacobs, and Chad~R. Sprouse.
\newblock Preconditioned quantum linear system algorithm.
\newblock \emph{Physical Review Letters}, 110:\penalty0 250504, 2013.
\newblock \doi{10.1103/PhysRevLett.110.250504}.

\bibitem[Bravo-Prieto et~al.(2023)Bravo-Prieto, LaRose, Cerezo, Suba\c{s}\i,
  Cincio, and Coles]{bravo2023variational}
Carlos Bravo-Prieto, Ryan LaRose, M.~Cerezo, Yi\u{g}it Suba\c{s}\i, Lukasz
  Cincio, and Patrick~J. Coles.
\newblock Variational quantum linear solver.
\newblock \emph{Quantum}, 7:\penalty0 1188, 2023.
\newblock \doi{10.22331/q-2023-11-22-1188}.

\bibitem[Huang et~al.(2019)Huang, Bharti, and Rebentrost]{huang2019near}
Hsin-Yuan Huang, Kishor Bharti, and Patrick Rebentrost.
\newblock Near-term quantum algorithms for linear systems of equations.
\newblock \emph{arXiv preprint arXiv:1909.07344}, 2019.

\bibitem[Lubasch et~al.(2020)Lubasch, Joo, Moinier, Kiffner, and
  Jaksch]{lubasch2020variational}
Michael Lubasch, Jae~Woo Joo, Pierre Moinier, Martin Kiffner, and Dieter
  Jaksch.
\newblock Variational quantum algorithms for nonlinear problems.
\newblock \emph{Physical Review A}, 101:\penalty0 010301(R), 2020.
\newblock \doi{10.1103/PhysRevA.101.010301}.

\bibitem[Sarma et~al.(2024)Sarma, Watts, Moosa, Liu, and
  McMahon]{sarma2024quantum}
Abhijat Sarma, Thomas~W. Watts, Mudassir Moosa, Yilian Liu, and Peter~L.
  McMahon.
\newblock Quantum variational solving of nonlinear and multidimensional partial
  differential equations.
\newblock \emph{Physical Review A}, 109:\penalty0 062616, 2024.
\newblock \doi{10.1103/PhysRevA.109.062616}.

\bibitem[Ayoub and Baeder(2025)]{ayoub2025poisson}
Fouad Ayoub and James~D. Baeder.
\newblock High-entanglement capabilities for variational quantum algorithms:
  the {Poisson} equation case.
\newblock \emph{Quantum Information Processing}, 24:\penalty0 229, 2025.
\newblock \doi{10.1007/s11128-025-04846-y}.

\bibitem[Trefethen(2000)]{trefethen2000spectral}
Lloyd~N. Trefethen.
\newblock \emph{Spectral Methods in MATLAB}.
\newblock SIAM, Philadelphia, 2000.
\newblock \doi{10.1137/1.9780898719598}.

\bibitem[Boyd(2001)]{boyd2001chebyshev}
John~P. Boyd.
\newblock \emph{Chebyshev and Fourier Spectral Methods}.
\newblock Dover, Mineola, NY, 2 edition, 2001.

\bibitem[Canuto et~al.(2006)Canuto, Hussaini, Quarteroni, and
  Zang]{canuto2006spectral}
Claudio Canuto, M.~Yousuff Hussaini, Alfio Quarteroni, and Thomas~A. Zang.
\newblock \emph{Spectral Methods: Fundamentals in Single Domains}.
\newblock Springer, Berlin, 2006.
\newblock \doi{10.1007/978-3-540-30726-6}.

\bibitem[Shen et~al.(2011)Shen, Tang, and Wang]{shen2011spectral}
Jie Shen, Tao Tang, and Li-Lian Wang.
\newblock \emph{Spectral Methods: Algorithms, Analysis and Applications}.
\newblock Springer, Berlin, 2011.
\newblock \doi{10.1007/978-3-540-71041-7}.

\bibitem[Reed and Hill(1973)]{reed1973triangular}
William~H. Reed and T.~R. Hill.
\newblock Triangular mesh methods for the neutron transport equation.
\newblock \emph{Los Alamos Scientific Laboratory Report LA-UR-73-479}, 1973.

\bibitem[Bassi and Rebay(1997)]{bassi1997high}
Francesco Bassi and Stefano Rebay.
\newblock A high-order accurate discontinuous finite element method for the
  numerical solution of the compressible {Navier--Stokes} equations.
\newblock \emph{Journal of Computational Physics}, 131\penalty0 (2):\penalty0
  267--279, 1997.
\newblock \doi{10.1006/jcph.1996.5572}.

\bibitem[Cockburn and Shu(1998)]{cockburn1998local}
Bernardo Cockburn and Chi-Wang Shu.
\newblock The local discontinuous {Galerkin} method for time-dependent
  convection-diffusion systems.
\newblock \emph{SIAM Journal on Numerical Analysis}, 35\penalty0 (6):\penalty0
  2440--2463, 1998.
\newblock \doi{10.1137/S0036142997316712}.

\bibitem[Cockburn et~al.(2000)Cockburn, Karniadakis, and
  Shu]{cockburn2001runge}
Bernardo Cockburn, George~E. Karniadakis, and Chi-Wang Shu.
\newblock The development of discontinuous {Galerkin} methods.
\newblock \emph{Lecture Notes in Computational Science and Engineering},
  11:\penalty0 3--50, 2000.

\bibitem[Arnold et~al.(2002)Arnold, Brezzi, Cockburn, and
  Marini]{arnold2002unified}
Douglas~N. Arnold, Franco Brezzi, Bernardo Cockburn, and L.~Donatella Marini.
\newblock Unified analysis of discontinuous {Galerkin} methods for elliptic
  problems.
\newblock \emph{SIAM Journal on Numerical Analysis}, 39\penalty0 (5):\penalty0
  1749--1779, 2002.
\newblock \doi{10.1137/S0036142901384162}.

\bibitem[Rivi{\`e}re(2008)]{riviere2008discontinuous}
Beatrice Rivi{\`e}re.
\newblock \emph{Discontinuous Galerkin Methods for Solving Elliptic and
  Parabolic Equations}.
\newblock SIAM, Philadelphia, 2008.
\newblock \doi{10.1137/1.9780898717440}.

\bibitem[Hesthaven and Warburton(2008)]{hesthaven2008nodal}
Jan~S. Hesthaven and Tim Warburton.
\newblock \emph{Nodal Discontinuous Galerkin Methods: Algorithms, Analysis, and
  Applications}.
\newblock Springer, New York, 2008.
\newblock \doi{10.1007/978-0-387-72067-8}.

\bibitem[Di~Pietro and Ern(2012)]{dipietro2012mathematical}
Daniele~A. Di~Pietro and Alexandre Ern.
\newblock \emph{Mathematical Aspects of Discontinuous Galerkin Methods}.
\newblock Springer, Berlin, 2012.
\newblock \doi{10.1007/978-3-642-22980-0}.

\bibitem[Nitsche(1971)]{nitsche1971}
Joachim Nitsche.
\newblock {\"U}ber ein variationsprinzip zur l{\"o}sung von
  {Dirichlet}-problemen bei verwendung von teilr{\"a}umen, die keinen
  randbedingungen unterworfen sind.
\newblock \emph{Abhandlungen aus dem Mathematischen Seminar der Universit{\"a}t
  Hamburg}, 36:\penalty0 9--15, 1971.
\newblock \doi{10.1007/BF02995904}.

\bibitem[Dolean et~al.(2015)Dolean, Jolivet, and Nataf]{dolean2015domain}
Victor Dolean, Pierre Jolivet, and Fr{\'e}d{\'e}ric Nataf.
\newblock \emph{An Introduction to Domain Decomposition Methods: Algorithms,
  Theory, and Parallel Implementation}.
\newblock SIAM, Philadelphia, 2015.
\newblock \doi{10.1137/1.9781611974065}.

\bibitem[LeVeque(2007)]{leveque2007finite}
Randall~J. LeVeque.
\newblock \emph{Finite Difference Methods for Ordinary and Partial Differential
  Equations}.
\newblock SIAM, Philadelphia, 2007.
\newblock \doi{10.1137/1.9780898717839}.

\bibitem[Strikwerda(2004)]{strikwerda2004finite}
John~C. Strikwerda.
\newblock \emph{Finite Difference Schemes and Partial Differential Equations}.
\newblock SIAM, Philadelphia, 2 edition, 2004.
\newblock \doi{10.1137/1.9780898717938}.

\bibitem[Grover and Rudolph(2002)]{grover2002creating}
Lov~K. Grover and Terry Rudolph.
\newblock Creating superpositions that correspond to efficiently integrable
  probability distributions.
\newblock \emph{arXiv preprint quant-ph/0208112}, 2002.

\bibitem[M{\"o}tt{\"o}nen et~al.(2005)M{\"o}tt{\"o}nen, Vartiainen, Bergholm,
  and Salomaa]{mottonen2005transformation}
Mikko M{\"o}tt{\"o}nen, Juha~J. Vartiainen, Ville Bergholm, and Martti~M.
  Salomaa.
\newblock Transformation of quantum states using uniformly controlled
  rotations.
\newblock \emph{Quantum Information \& Computation}, 5\penalty0 (6):\penalty0
  467--473, 2005.

\bibitem[Shende et~al.(2006)Shende, Bullock, and Markov]{shende2006synthesis}
Vivek~V. Shende, Stephen~S. Bullock, and Igor~L. Markov.
\newblock Synthesis of quantum-logic circuits.
\newblock \emph{IEEE Transactions on Computer-Aided Design of Integrated
  Circuits and Systems}, 25\penalty0 (6):\penalty0 1000--1010, 2006.
\newblock \doi{10.1109/TCAD.2005.855930}.

\bibitem[Plesch and Brukner(2011)]{plesch2011quantum}
Martin Plesch and {\v C}aslav Brukner.
\newblock Quantum-state preparation with universal gate decompositions.
\newblock \emph{Physical Review A}, 83:\penalty0 032302, 2011.
\newblock \doi{10.1103/PhysRevA.83.032302}.

\bibitem[Moosa et~al.(2023)Moosa, Watts, Chen, Sarma, and
  McMahon]{moosa2023linear}
Mudassir Moosa, Thomas~W. Watts, Yiyou Chen, Abhijat Sarma, and Peter~L.
  McMahon.
\newblock Linear-depth quantum circuits for loading fourier approximations of
  arbitrary functions.
\newblock \emph{arXiv preprint arXiv:2302.03888}, 2023.

\bibitem[Quarteroni et~al.(2007)Quarteroni, Sacco, and
  Saleri]{quarteroni2007numerical}
Alfio Quarteroni, Riccardo Sacco, and Fausto Saleri.
\newblock \emph{Numerical Mathematics}.
\newblock Springer, Berlin, 2 edition, 2007.
\newblock \doi{10.1007/b98885}.

\bibitem[Ern and Guermond(2004)]{ern2004theory}
Alexandre Ern and Jean-Luc Guermond.
\newblock \emph{Theory and Practice of Finite Elements}.
\newblock Springer, New York, 2004.
\newblock \doi{10.1007/978-1-4757-4355-5}.

\bibitem[Tang(2005)]{tang2005moving}
Tao Tang.
\newblock Moving mesh methods for computational fluid dynamics.
\newblock In \emph{Recent Advances in Adaptive Computation}, pages 141--173.
  American Mathematical Society, Providence, RI, 2005.

\bibitem[Budd et~al.(2009)Budd, Huang, and Russell]{budd2009adaptivity}
Chris~J. Budd, Weizhang Huang, and Robert~D. Russell.
\newblock Adaptivity with moving grids.
\newblock \emph{Acta Numerica}, 18:\penalty0 111--241, 2009.
\newblock \doi{10.1017/S0962492906400015}.

\bibitem[Zhu et~al.(2024)Zhu, Liang, Yang, and Li]{zhu2024optimizing}
Linghua Zhu, Senwei Liang, Chao Yang, and Xiaosong Li.
\newblock Optimizing shot assignment in variational quantum eigensolver
  measurement.
\newblock \emph{Journal of Chemical Theory and Computation}, 20:\penalty0
  10.1021/acs.jctc.3c01113, 2024.
\newblock \doi{10.1021/acs.jctc.3c01113}.

\bibitem[Nakaji et~al.(2023)]{nakaji2023measurement}
Kouhei Nakaji et~al.
\newblock Measurement optimization of variational quantum simulation by
  classical shadow and derandomization.
\newblock \emph{Quantum}, 7:\penalty0 995, 2023.
\newblock \doi{10.22331/q-2023-05-01-995}.

\end{thebibliography}
\end{document}